\documentclass{LMCS}

\def\dOi{9(4:16)2013}
\lmcsheading%
{\dOi}
{1--34}
{}
{}
{Jan.~13, 2013}
{Dec.~\phantom04, 2013}
{}

\ACMCCS{[{\bf Mathematics of computing}]: Probability and
  statistics---Stochastic processes---Markov processes; [{\bf Theory
      of computation}]: Formal languages and automata
  theory---Automata over infinite objects}

\pdfoutput=1 

\usepackage[utf8]{inputenc}
\usepackage[T1]{fontenc}
\usepackage[english]{babel}				
\usepackage{amsmath, amssymb}		
\usepackage{lmodern}	

\usepackage{hyperref,ifthen}

\usepackage[arrow,matrix,curve,color,all]{xy} 	
\UseComputerModernTips 				
\usepackage{tikz}
\usetikzlibrary{arrows, positioning, automata, matrix}
				
\usepackage{xargs}						 

\newcommand{\Meas}{\mathbf{Meas}}
\newcommand{\SeT}{\mathbf{Set}}
\newcommand{\one}{\mathbf{1}}

\newcommand{\Id}{\mathrm{Id}}
\newcommand{\id}{\mathrm{id}}
\newcommand{\Kl}{\mathcal{K}\ell}

\newcommand{\G}{\mathcal{G}}
\newcommand{\powerset}[1]{\mathcal{P}\left({#1}\right)}
\newcommand{\set}[1]{\left\{#1\right\}}

\newcommand{\tr}{\mathbf{tr}}

\renewcommand{\epsilon}{\varepsilon}
\renewcommand{\phi}{\varphi}

\newcommand{\A}{\ensuremath{\mathcal{A}}}
\newcommand{\Astar}{{\mathcal{A}^*}}
\newcommand{\Aomega}{{\mathcal{A}^\omega}}
\newcommand{\Ainfty}{{\mathcal{A}^\infty}}

\renewcommand{\S}{\mathcal{S}}
\newcommand{\Sstar}{{\mathcal{S}_*}}
\newcommand{\Somega}{{\mathcal{S}_\omega}}
\newcommand{\Sinfty}{{\mathcal{S}_\infty}}

\renewcommand{\G}{\mathcal{G}}

\newcommand{\sigalg}[2][]{\sigma_{#1}(#2)}

\newcommand{\N}{\mathbb{N}}

\newcommand{\R}{\mathbb{R}}
\newcommand{\ExtR}{\overline{\mathbb{R}}}
\newcommand{\cone}[2]{\uparrow_{#1}\!\{#2\}}

\newcommand{\Borel}{\mathcal{B}}
\newcommand{\Lebesgue}{\mathcal{L}}

\newcommand{\Prob}[1]{\mathbb{P}\left({#1}\right)}

\newcommand{\Intersections}[1]{\ensuremath{\mathcal{I}\left(#1\right)}}
\newcommand{\Unions}[1]{\ensuremath{\mathcal{U}\left(#1\right)}}

\newcommand{\Rdiamond}{\mathcal{R}_\diamond}
\newcommand{\compl}[1]{\ensuremath{\A^\diamond \setminus #1}}
\renewcommand{\cone}[2]{\ensuremath{#2\A^{#1}}}

\renewcommand{\P}[3]{\ensuremath{\mathbf{P}_{#1}\left(#2, #3\right)}}

\newcommandx{\Int}[5][1=\empty,2=\empty,4=x,5=\empty]{%
	\ifthenelse{\equal{#5}{\empty}}%
	{%
		\ifthenelse{\equal{#1}{\empty}}%
			{\int\!{#3}\,\mathrm{d}{#4}}%
			{\ifthenelse{\equal{#2}{\empty}}%
				{\int_{#1}\!{#3}\,\mathrm{d}{#4}}%
				{\int_{#1}^{#2}\!#3\,\mathrm{d}{#4}}%
			}%
	}%
	{%
		\ifthenelse{\equal{#1}{\empty}}%
			{\int#5\!{#3}\,\mathrm{d}{#4}}%
			{\ifthenelse{\equal{#2}{\empty}}%
				{\int#5_{#1}\!{#3}\,\mathrm{d}{#4}}%
				{\int#5_{#1}^{#2}\!#3\,\mathrm{d}{#4}}%
			}%
	}%
}%

\begin{document}

\title[Coalgebraic Trace Semantics for Continuous PTS]{Coalgebraic Trace Semantics for Continuous Probabilistic Transition Systems\rsuper*}

\author[Henning Kerstan]{Henning Kerstan}	
\address{Universität Duisburg-Essen, Duisburg, Germany}	
\email{\{henning.kerstan, barbara\_koenig\}@uni-due.de}  

\author[Barbara König]{Barbara König}

\titlecomment{{\lsuper*}This is an extended version of \cite{KK12a} which was presented at CONCUR 2012.}

\hypersetup{%
	pdftitle={Coalgebraic Trace Semantics for Continuous Probabilistic Transition Systems},%
	pdfauthor={Henning Kerstan, Barbara König},%
	pdfkeywords={probabilistic transition systems, Markov processes, coalgebra, trace semantics}
}

\begin{abstract}
Coalgebras in a Kleisli category yield a generic definition of trace semantics for various types of labelled transition systems. In this paper we apply this generic theory to generative probabilistic transition systems, short PTS, with arbitrary (possibly uncountable) state spaces. We consider the sub-probability monad and the probability monad (Giry monad) on the category of measurable spaces and measurable functions. Our main contribution is that the existence of a final coalgebra in the Kleisli category of these monads is closely connected to the measure-theoretic extension theorem for sigma-finite pre-measures. In fact, we obtain a practical definition of the trace measure for both finite and infinite traces of PTS that subsumes a well-known result for discrete probabilistic transition systems. Finally we consider two example systems with uncountable state spaces and apply our theory to calculate their trace measures. 
\end{abstract}

\maketitle

\section{Introduction}
Coalgebra \cite{jacobs,r:universal-coalgebra} is a general framework in which several types of transition systems can be studied (deterministic and non-deterministic automata, weighted automata, transition systems with non deterministic and probabilistic branching, etc.). One of the strong points of coalgebra is that it induces -- via the notion of coalgebra homomorphism and final coalgebra -- a notion of behavioral equivalence for all these types of systems. The resulting behavioral equivalence is usually some form of bisimilarity.  However, \cite{hasuo} has shown that by modifying the category in which the coalgebra lives, one can obtain different notions of behavioral equivalence, such as trace equivalence.

We will shortly describe the basic idea: given an endofunctor $F$ on $\SeT$, the category of sets and total functions, describing the branching type of the system, a coalgebra in the category $\mathbf{Set}$ is a function $\alpha\colon X\to FX$, where $X$ is a set. Consider, for instance, the functor $FX = \mathcal{P}_\mathit{fin}(\A \times X+\one)$, where $\mathcal{P}_\mathit{fin}$ is the finite powerset functor and $\A$ is a given alphabet. This setup allows us to specify finitely branching non-deterministic automata where a state $x\in X$ is mapped to a set of tuples of the form $(a,y)$, for $a\in \A, y\in X$, describing transitions. The set contains the symbol $\checkmark$ (for termination) -- the only element contained in the one-element set $\one$ -- if and only if $x$ is a final state.

A coalgebra homomorphism maps the set of states of a coalgebra to the set of states of another coalgebra, preserving the branching structure. Furthermore, the final coalgebra -- if it exists -- is the final object in the category of coalgebras. Every coalgebra has a unique homomorphism into the final coalgebra and two states of a transition system modelled as coalgebra are mapped to the same state in the final coalgebra iff they are behaviorally equivalent.

Now, applying this notion to the example above induces bisimilarity, whereas usually the appropriate notion of behavioral equivalence for non-deterministic finite automata is language equivalence. One of the ideas of \cite{hasuo} is to view a coalgebra $X\to\mathcal{P}(\mathcal{A}\times X+\one)$ not as an arrow in $\mathbf{Set}$, but as an arrow $X\to \mathcal{A}\times X+\one$ in $\mathbf{Rel}$, the category of sets and relations which is also the Kleisli category of the powerset monad. This induces trace equivalence, instead of bisimilarity, with the underlying intuition that non-determinism is a side-effect that is ``hidden'' within the monad. This side effect is not present in the final coalgebra (which consists of the set $\mathcal{A}^*$ with a suitable coalgebra structure), but in the arrow from a state $x\in X$ to $\mathcal{A}^*$, which is a relation, and relates each state with all words accepted from this state.

More generally, coalgebras are given as arrows $X\to TFX$ in a Kleisli category, where a monad $T$ describes implicit branching and an endofunctor $F$ specifies explicit branching with the underlying intuition that the implicit branching (for instance non-determinism or probabilistic branching) is aggregated and abstracted away in the final coalgebra. For several monads this yields a form of trace semantics. In \cite{hasuo} a theorem gives sufficient conditions for the existence of a final coalgebra for Kleisli categories over $\mathbf{Set}$, which -- interestingly -- can be obtained as initial $F$-algebra in $\mathbf{Set}$.

In \cite{hasuo} it is also proposed to obtain probabilistic trace semantics for the Kleisli category of the (discrete) subdistribution monad $\mathcal{D}$ on $\SeT$. The endofunctor of that monad maps a set $X$ to the set $\mathcal{D}(X)$ of all functions $p\colon X \to [0,1]$ satisfying $\sum_{x \in X} p(x) \leq 1$. Coalgebras in this setting are functions of the form $X\to \mathcal{D}(\mathcal{A}\times X+\one)$ (modeling probabilistic branching and termination), seen as arrows in the corresponding Kleisli category. From the general result in \cite{hasuo} mentioned above it again follows that the final coalgebra is carried by $\mathcal{A}^*$, where the mapping into the final coalgebra assigns to each state a discrete probability distribution over its traces. In this way one obtains the finite trace semantics of generative probabilistic systems \cite{s:coalg-ps-phd,glabbeek}.

The contribution in \cite{hasuo} is restricted to discrete probability spaces, where the probability distributions always have at most countable support \cite{Sokolova20115095}. This might seem sufficient for practical applications at first glance, but it has two important drawbacks: first, it excludes several interesting systems that involve uncountable state spaces (see for instance the examples in Section~\ref{sec:advexamples} or the examples in \cite{Pan09}). Second, it excludes the treatment of infinite traces, as detailed in \cite{hasuo}, since the set of all infinite traces is uncountable and hence needs measure theory to be treated appropriately. This is an intuitive reason for the choice of the subdistribution monad -- instead of the distribution monad -- in \cite{hasuo}: for a given state, it might always be the case that a non-zero ``probability mass'' is associated to the infinite traces leaving this state, which -- in the discrete case -- cannot be specified by a probability distribution over all words.

Hence, we generalize the results concerning probabilistic trace semantics from \cite{hasuo} to the case of uncountable state spaces, by working in the Kleisli category of the (continuous) sub-probability monad over $\Meas$ (the category of measurable spaces). Unlike in \cite{hasuo} we do not derive the final coalgebra via a generic construction (building the initial algebra of the functor), but we construct the final coalgebra directly. Furthermore we consider the Kleisli category of the (continuous) probability monad (Giry monad) and treat the case with and without termination. In the former case we obtain a coalgebra over the set $\mathcal{A}^\infty$ (finite and infinite traces over $\mathcal{A}$) and in the latter over the set $\mathcal{A}^\omega$ (infinite traces), which shows the naturality of the approach. For completeness we also consider the case of the sub-probability monad without termination, which results in a trivial final coalgebra over the empty set.  In all cases we obtain the natural trace measures as instances of the generic coalgebraic theory.

Since, to our knowledge, there is no generic construction of the final coalgebra for these cases, we construct the respective final coalgebras directly and show their correctness by proving that each coalgebra admits a unique homomorphism into the final coalgebra. Here we rely on the measure-theoretic extension theorem for sigma-finite pre-measures and the identity theorem.

In the conclusion we will further compare our approach to \cite{hasuo} and discuss why we took an alternative route.

\subsection{Another paper?}
This paper is the extended version of the paper \cite{KK12a} first published at CONCUR 2012 and thus it necessarily contains all results of that paper. Due to page limitations some of the proofs were omitted in the published version and hence in the technical report \cite{KK12TR} we provided a version which is identical to the original paper but contains an appendix with the missing proofs. In contrast to that, the paper at hand contains all the proofs in place and also some corrections. Moreover, more details are presented, mainly taken from \cite{kerstan}, which was the starting point for everything. Last but not least the paper at hand includes the new Section \ref{sec:advexamples} containing two examples with uncountable state spaces and some additional theory needed in order to understand them.

\section{Background Material and Preliminaries}
\label{sec:prelim}
We assume that the reader is familiar with the basic definitions of category theory. However, we will provide a brief introduction to notation, measure theory and integration, coalgebra, coalgebraic trace semantics and Kleisli categories -- of course all geared to our needs. 

\subsection{Notation}
By $\one$ we denote a singleton set, its unique element is $\checkmark$. For arbitrary sets $X, Y$ we write $X \setminus Y$ for set complement, $X \times Y$ for the usual cartesian product and the disjoint union $X + Y$ is the set $\set{(x,0), (y,1)\mid x \in X, y \in Y}$. Whenever $X \cap Y = \emptyset$ this coincides with (is isomorphic to) the usual union $X \cup Y$ in an obvious way. For set inclusion we write $\subset$ for strict inclusion and $\subseteq$ otherwise. The set of real numbers is denoted by $\R$, the set of extended reals is the set $\ExtR := \R \cup \set{\pm\infty}$ and $\R_+$ and $\ExtR_+$ are their restrictions to the non-negative (extended) reals. We require $0 \cdot \pm \infty = \pm \infty \cdot 0 = 0$. For a function $f\colon X \to Y$ and a set $A \subseteq X$ the restriction of $f$ to $A$ is the function $f|_A\colon A \to Y$.  

\subsection{A Brief Introduction to Measure Theory}
Within this section we want to give a very brief introduction to measure theory. For a more thorough treatment there are many standard textbooks as e.g. {\cite{ash,Els07}}. Measure theory generalizes the idea of length, area or volume. Its most basic definition is that of a $\emph{$\sigma$-algebra (sigma-algebra)}$. Given an arbitrary set $X$ we call a set $\Sigma$ of subsets of $X$ a \emph{$\sigma$-algebra} iff it contains the empty set and is closed under complement and countable union. The tuple $(X, \Sigma)$ is called a \emph{measurable space}. We will sometimes call the set $X$ itself a measurable space, keeping in mind that there is an associated $\sigma$-algebra which we will then denote by $\Sigma_X$. For any subset $\mathcal{G} \subseteq \powerset{X}$ we can always uniquely construct the smallest $\sigma$-algebra on $X$ containing $\mathcal{G}$ which is denoted by $\sigalg[X]{\mathcal{G}}$. We call $\mathcal{G}$ the \emph{generator} of $\sigalg[X]{\mathcal{G}}$, which in turn is called \emph{the $\sigma$-algebra generated by $\mathcal{G}$}. It is known (and easy to show), that $\sigma_X$ is a monotone and idempotent operator. The elements of a $\sigma$-algebra on $X$ are called the \emph{measurable sets} of $X$. Among all possible generators for $\sigma$-algebras, there are special ones, so-called \emph{semirings of sets}.

\begin{defi}[Semiring of Sets]
Let $X$ be an arbitrary set. A subset $\S \subseteq \powerset{X}$ is called a \emph{semiring of sets} if it satisfies the following three properties.
\begin{enumerate}[label=(\alph*)]
	\item $\S$ contains the empty set, i.e. $\emptyset \in \S$.
	\item $\S$ is closed under pairwise intersection, i.e. for $A, B \in \S$ we always require $(A \cap B)\in \S$.
	\item The set difference of any two sets in $\S$ is the disjoint union of finitely many sets in $\S$, i.e. for any $A, B \in \S$ there is an $N \in \N$ and pairwise disjoint sets $C_1,\hdots,C_N \in \S$ such that $A\setminus B = \cup_{n=1}^N C_n$.
\end{enumerate}
\end{defi}

\noindent It is easy to see that every $\sigma$-algebra is a semiring of sets but the reverse is false. Please note that a semiring of sets is different from a semiring in algebra. For our purposes, we will consider special semirings containing a countable cover of the base set. 

\begin{defi}[Countable Cover, Covering Semiring]
Let $\S$ be a semiring. A countable sequence $(S_n)_{n \in \N}$ of sets in $\S$ such that $\cup_{n \in \N}S_n = X$ is called a \emph{countable cover of $X$ (in $\S$)}. If such a countable cover exists we call $\S$ a \emph{covering} semiring.
\end{defi}

With these basic structures at hand, we can now define pre-measures and measures. A non-negative function $\mu \colon \S \to \ExtR_+$ defined on a semiring $\S$ is called a \emph{pre-measure} on $X$ if it assigns $0$ to the empty set and is \emph{$\sigma$-additive}, i.e. for a sequence $(S_n)_{n \in \N}$ of pairwise disjoint sets in $\mathcal{S}$ where $\left(\cup_{n \in \N}S_n\right) \in \mathcal{S}$ we must have 
\begin{align}
	\mu\left(\bigcup_{n \in \N}S_n\right) = \sum_{n \in \N}\mu\left(S_n\right).
\end{align}  
A pre-measure $\mu$ is called  \emph{$\sigma$-finite} if there is a countable cover $(S_n)_{n \in \N}$ of $X$ in $\mathcal{S}$ such that $\mu\left(S_n\right) < \infty$ for all $n \in \N$. Whenever $\S$ is a $\sigma$-algebra we call $\mu$ a \emph{measure} and the tuple $(X, \S, \mu)$ a \emph{measure space}. In that case $\mu$ is said to be \emph{finite} iff $\mu(X) < \infty$ and for the special cases $\mu(X) = 1$ (or $\mu(X) \leq 1$) $\mu$ is called a \emph{probability measure} (or \emph{sub-probability measure} respectively). Measures are \emph{monotone}, i.e. if $A,B$ are measurable $A \subseteq B$ implies $\mu(A) \leq \mu(B)$ and \emph{continuous}, i.e. for measurable $A_1 \subseteq A_2 \subseteq \hdots \subseteq A_n \subseteq \hdots$ we always have $\mu\left(\cup_{n=1}^\infty A_n\right) = \lim_{n \to \infty} \mu(A_n)$ and for measurable $B_1 \supseteq B_2 \supseteq \hdots \supseteq B_n \supseteq \hdots$ with $\mu(B_1) < \infty$ we have  $\mu\left(\cap_{n=1}^\infty A_n\right) = \lim_{n \to \infty} \mu(A_n)$ \cite[1.2.5 and 1.2.7]{ash}.

Given a measurable space $(X, \Sigma_X)$, a simple and well-known probability measure, is the so-called \emph{Dirac measure}, which we will use later. It is defined as $\delta_x^X\colon \Sigma_X \to [0,1]$, and is $1$ on $S \in \Sigma_X$ iff $x \in S$ and $0$ otherwise.

The most significant theorems from measure theory which we will use in this paper are the identity theorem and the extension theorem for $\sigma$-finite pre-measures, for which a proof can be found e.g. in~\cite[II.5.6 and II.5.7]{Els07}.

\begin{prop}[Identity Theorem]
Let $X$ be a set, $\G \subseteq \powerset{X}$ be a set which is closed under pairwise intersection and $\mu, \nu \colon \sigalg[X]{\G} \to \ExtR_+$ be measures. If $\mu|_\G = \nu|_\G$ and $\G$ contains a countable cover $(G_n)_{n \in \N}$ of $X$ satisfying $\mu(G_n) = \nu(G_n) < \infty$ for all $n \in \N$ then $\mu = \nu$.\qed
\end{prop}

\begin{prop}[Extension Theorem for $\sigma$-finite Pre-Measures]
\label{prop:extension}
Let $X$ be a set, \linebreak $\S \subseteq \powerset{X}$ be a semiring of sets and $\mu\colon \S \to \ExtR_+$ be a $\sigma$-finite pre-measure. Then there exists a uniquely determined measure $\hat{\mu} \colon \sigalg[X]{\S} \to \ExtR_+$ such that $\hat{\mu}|_\S = \mu$. \qed
\end{prop} 

As we are only interested in finite measures, we provide a result, which can be derived easily from the identity theorem. 

\begin{cor}[Equality of Finite Measures on Covering Semirings]
\label{cor:equality_of_measures}
Let $X$ be an arbitrary set, $\S \subseteq \powerset{X}$ be a covering semiring and $\mu, \nu \colon \sigalg[X]{\S} \to \ExtR_+$ be finite measures. Then $\mu = \nu$  if and only if $\mu|_\S = \nu|_\S$.
\end{cor}
\proof
Obviously we get $\mu|_\S = \nu|_\S$ if $\mu = \nu$. For the other direction let $(S_n)_{n \in \N}$ be a countable cover of $X$. Then finiteness of $\mu$ and $\nu$ together with the fact that measures are continuous and $\mu|_\S = \nu|_\S$ yield $\mu(S_n) = \nu(S_n) \leq \nu(X)< \infty$ for all $n \in \N$. Since $\S$ is a semiring of sets, it is closed under pairwise intersection which allows us to apply the identity theorem yielding $\mu = \nu$. 
\qed

\subsection{The Category of Measurable Spaces and Functions}
Let $X$ and $Y$ be measurable spaces. A function $f \colon X \to Y$ is called \emph{measurable} iff the pre-image of any measurable set of $Y$ is a measurable set of $X$. The category $\Meas$ has measurable spaces as objects and measurable functions as arrows.  Composition of arrows is function composition and the identity arrows are the identity functions.

The product of two measurable spaces $(X, \Sigma_X)$ and $(Y, \Sigma_Y)$  is the set $X \times Y$ endowed with the $\sigma$-algebra generated by $\Sigma_X \ast \Sigma_Y$, the set of so-called ``rectangles'' of measurable sets which is $\set{S_X \times S_Y\mid S_X \in \Sigma_X, S_Y \in \Sigma_Y}$. It is called the \emph{product $\sigma$-algebra} of $\Sigma_X$ and $\Sigma_Y$ and is denoted by $\Sigma_X \otimes \Sigma_Y$. Whenever $\Sigma_X$ and $\Sigma_Y$ have suitable generators, we can also construct a possibly smaller generator for the product $\sigma$-algebra by taking only the ``rectangles'' of the generators.

\begin{prop}[Generators for the Product $\sigma$-Algebra]
\label{prop:generator_product}
Let $X, Y$ be arbitrary sets and $\G_X \subseteq \powerset{X}, \G_Y \subseteq \powerset{Y}$ such that $X \in \G_X$ and $Y \in \G_Y$. Then the following holds: 
\[
	\sigalg[X\times Y]{\G_X \ast \G_Y} = \sigalg[X]{\G_X} \otimes \sigalg[Y]{\G_Y}\,.  \eqno{\qEd}
\]
\end{prop}

A proof of this proposition can be found in many standard textbooks on measure theory, e.g. in \cite{Els07}. We remark that there are (obvious) product endofunctors on the category of measurable spaces and functions.

\begin{defi}[Product Functors]
Let $(Z, \Sigma_Z)$ be a measurable space. The endofunctor $Z \times \Id_\Meas$ maps a measurable space $(X, \Sigma_X)$ to $\left(Z \times X, \Sigma_Z \otimes \Sigma_X\right)$ and a measurable function $f\colon X \to Y$ to the measurable function $Z \times f \colon Z \times X \to Z \times Y, (z,x) \mapsto \left(z,f(x)\right)$. The functor $\Id_\Meas \times Z$ is constructed analogously.
\end{defi}

The coproduct of two measurable spaces $(X, \Sigma_X)$ and $(Y, \Sigma_Y)$ is the set $X + Y$ endowed with $\Sigma_X \oplus \Sigma_Y := \set{S_X + S_Y\mid S_X \in \Sigma_X, S_Y \in \Sigma_Y}$ as $\sigma$-algebra, the \emph{disjoint union $\sigma$-algebra}. Note that in contrast to the product no $\sigma$-operator is needed because $\Sigma_X \oplus \Sigma_Y$ itself is already a $\sigma$-algebra whereas $\Sigma_X \ast \Sigma_Y$ is usually no $\sigma$-algebra. For generators of the disjoint union $\sigma$-algebra we provide and prove a comparable result to the one given above for the product $\sigma$-algebra. 
\begin{prop}[Generators for the Disjoint Union $\sigma$-Algebra]
\label{prop:generator_union}
Let $X, Y$ be arbitrary sets and $\G_X \subseteq \powerset{X}, \G_Y \subseteq \powerset{Y}$ such that $\emptyset \in \G_X$ and $Y \in \G_Y$. Then the following holds:
\begin{align}
 	\sigalg[X + Y]{\G_X \oplus \G_Y} = \sigalg[X]{\G_X} \oplus \sigalg[Y]{\G_Y}\label{eq:generator_union}\,.
\end{align}
\end{prop}
\noindent In order to prove this, we cite another result from \cite[I.4.5 Korollar]{Els07}.

\begin{lem}
\label{lem:trace_sigma_algebra}
Let $X$ be an arbitrary set, $\mathcal{G} \subseteq \powerset{X}$ and $S \subseteq X$. Then $\sigalg[S]{\mathcal{G}|S} = \sigma_X(\mathcal{G})|S$ where $\mathcal{G} | S := \set{G \cap S \mid G \in \mathcal{G}}$ and analogously $\sigalg[X]{\mathcal{G}}|S := \set{G \cap S \mid G \in \sigalg[X]{\mathcal{G}}}$. 
\end{lem}

\proof[Proof of Proposition~\ref{prop:generator_union}]
Without loss of generality we assume that $X$ and $Y$ are pairwise disjoint. Hence for any subsets $A \subseteq X$, $B \subseteq Y$ we have $A \cap B = \emptyset$ and thus $A + B \cong A \cup B$. In order to prove equation \eqref{eq:generator_union} we show both inclusions.
\begin{itemize}[label=$\subseteq$]
	\item We have $\mathcal{G}_X \oplus \mathcal{G}_Y \subseteq  \sigalg[X]{\mathcal{G}_X} \oplus \sigalg[Y]{\mathcal{G}_Y}$ and thus monotonicity and idempotence of the $\sigma$-operator immediately yield $\sigalg[X\cup Y]{\mathcal{G}_X \oplus \mathcal{G}_Y} \subseteq  \sigalg[X]{\mathcal{G}_X} \oplus \sigalg[Y]{\mathcal{G}_Y}$.
	\item [$\supseteq$] Let $G \in \sigalg[X]{\mathcal{G}_X} \oplus \sigalg[Y]{\mathcal{G}_Y}$. Then $G = G_X\cup G_Y$ with $G_X \in \sigalg[X]{\mathcal{G}_X}$ and $G_Y \in \sigalg[Y]{\mathcal{G}_Y}$. \linebreak We observe that $\mathcal{G}_X = (\mathcal{G}_X \oplus \mathcal{G}_Y) | X$ and by applying Lemma~\ref{lem:trace_sigma_algebra} we obtain that $\sigalg[X\cup Y]{\mathcal{G}_X \oplus \mathcal{G}_Y} | X = \sigalg[X]{\mathcal{G}_X}$. Thus there must be a $G'_Y \in \powerset{Y}$ such that \linebreak $G_X \cup G'_Y \in \sigalg[X\cup Y]{\mathcal{G}_X \oplus \mathcal{G}_Y}$. Analogously there must be a $G'_X \in \powerset{X}$  such that $G'_X \cup G_Y \in \sigalg[X\cup Y]{\mathcal{G}_X \oplus \mathcal{G}_Y}$. 
We have $Y = \emptyset \cup Y \in \sigalg[X\cup Y]{\mathcal{G}_X \oplus \mathcal{G}_Y}$ and hence we also have $X = (X\cup Y)\setminus Y  \in \sigalg[X\cup Y]{\mathcal{G}_X \oplus \mathcal{G}_Y}$. Thus we calculate 
\begin{align*}
	G = G_X \cup G_Y = \big( (G_X \cup G'_Y) \cap X \big) \cup \big( (G'_X \cup G_Y) \cap Y\big) \in \sigalg[X\cup Y]{\mathcal{G}_X \oplus \mathcal{G}_Y}
\end{align*}
and hence can conclude that $\sigalg[X\cup Y]{\mathcal{G}_X \oplus \mathcal{G}_Y} \supseteq  \sigalg[X]{\mathcal{G}_X} \oplus \sigalg[Y]{\mathcal{G}_Y}$.\qed
\end{itemize}

\noindent As before we have endofunctors for the coproduct, the coproduct functors.

\begin{defi}[Co-Product Functors]
Let $(Z, \Sigma_Z)$ be a measurable space. The endofunctor $\Id_\Meas + Z$ maps a measurable space $(X, \Sigma_X)$ to $\left(X+Z, \Sigma_X \oplus \Sigma_Z\right)$ and a measurable function $f\colon X \to Y$ to the measurable function $f + Z \colon X + Z\to Y + Z$, $(x,0) \mapsto (f(x),0)$, $(z,1) \mapsto (z,1)$. The functor $\Id_\Meas + Z$ is constructed analogously.
\end{defi}

For isomorphisms in $\Meas$ we provide the following characterization which we will need later for our main result.

\begin{prop}[Isomorphisms in $\Meas$]
\label{prop:isomorphisms}
Two measurable spaces $X$ and $Y$ are isomorphic in $\Meas$ iff there is a bijective function $\phi\colon X \to Y$ such that\footnote{For $\S \subseteq \powerset{X}$ and a function $\phi \colon X \to Y$ let $\phi(\S) = \set{\phi\left(S_X\right) \mid S_X \in \S} = \set{ \set{\phi(x) \mid x \in S_X} \mid S_X \in \S}$.} $\phi\left(\Sigma_X\right) = \Sigma_Y$. If $\Sigma_X$ is generated by a set $\S \subseteq \powerset{X}$ then $X$ and $Y$ are isomorphic iff there is a bijective function $\phi\colon X \to Y$ such that $\Sigma_Y$ is generated by ${\phi\left(\S\right)}$. In this case $\S$ is a (covering) semiring of sets [a $\sigma$-algebra] iff $\phi(\S)$ is a (covering) semiring of sets [a $\sigma$-algebra].
\end{prop}

Again, we need a result from measure theory for the proof. This auxiliary result and its proof can be found e.g. in \cite[I.4.4 Satz]{Els07}.

\begin{lem}
\label{lem:generator_inverse}
Let $X, Y$ be sets, $f\colon X \to Y$ be a function. Then for every subset $\S \subseteq \powerset{Y}$ it holds that $\sigalg[X]{f^{-1}(\S)} = f^{-1}\left(\sigalg[Y]{\S}\right)$.\qed
\end{lem}

\proof[Proof of Proposition~\ref{prop:isomorphisms}]
Since the identity arrows in $\Meas$ are the identity functions, we can immediately derive that any isomorphism $\phi\colon X \to Y$ must be a bijective function. Measurability of $\phi$ and its inverse function $\phi^{-1}\colon Y \to X$ yield $\phi\left(\Sigma_X\right) = \Sigma_Y$. The equality $\sigalg[Y]{\phi(\S)} = \phi\left(\sigalg[X]{\S}\right)$ follows from Lemma~\ref{lem:generator_inverse} by taking $f = \phi^{-1}$. The last equivalence is easy to verify using bijectivity of $\phi$ and $\phi^{-1}$.\qed

\subsection{Kleisli Categories and Liftings of Endofunctors}
Recall that a monad on a category $\mathbf{C}$ is a triple $(T, \eta, \mu)$ where $T\colon \mathbf{C} \to \mathbf{C}$ is an endofunctor together with two natural transformations\footnote{This is the second meaning of the symbol $\mu$. Until now, $\mu$ was used as a symbol for a (pre-)measure.} $\eta \colon \Id_\mathbf{C} \Rightarrow T$ and $\mu\colon T^2 \Rightarrow T$ such that the following diagrams commute for all $\mathbf{C}$-objects $X$.
\[\xymatrix@C+20 pt{
  TX
    \ar[r]^{T\eta_X}
    \ar[dr]_{\,\id_{TX}\!}
    \ar[d]_{\eta_{TX}}
& T^2X
    \ar[d]^{\mu_X}
&
& T^3X
    \ar[r]^{T\mu_X}
    \ar[d]_{\mu_{TX}}
& T^2X
    \ar[d]^{\mu_X}\\
  T^2X
    \ar[r]_{\mu_X}
& TX
&
& T^2X
    \ar[r]_{\mu_X}
& TX
}\]

\noindent Given a monad $(T, \eta, \mu)$ on a category $\mathbf{C}$ we can define a new category, the Kleisli category of $T$, where the objects are the same as in $\mathbf{C}$ but every arrow in the new category corresponds to an arrow $f\colon X \to TY$ in $\mathbf{C}$. Thus, arrows in the Kleisli category incorporate side effects specified by a monad~\cite{hasuo,abhkms:coalgebra-min-det}. Formally we will use the following definition. 

\begin{defi}[Kleisli Category]
Let $(T, \eta, \mu)$ be a monad on a category $\mathbf{C}$. The \emph{Kleisli category of $T$} has the same objects as $\mathbf{C}$. For any two such objects $X$ and $Y$, the Kleisli arrows with domain $X$ and codomain $Y$ are exactly the $\mathbf{C}$-arrows $f\colon X \to TY$. Composition of Kleisli arrows $f \colon X  \to TY$ and $g \colon Y\to TZ $ is defined as $g\circ_T f := \mu_Z \circ T(g)\circ f$, the identity arrow for any Kleisli object $X$ is $\eta_X$.
\end{defi}

Given an endofunctor $F$ on $\mathbf{C}$, we want to construct an endofunctor $\overline{F}$ on $\Kl(T)$ that ``resembles'' $F$: Since objects in $\mathbf{C}$ and objects in $\Kl(T)$ are the same, we want $\overline{F}$ to coincide with $F$ on objects i.e. we want $\overline{F}X = FX$. It remains to define how $\overline{F}$ shall act on Kleisli arrows $f\colon X \to TY$ such that it ``resembles'' $F$. Formally we require $\overline{F}$ to be a \emph{lifting} of $F$ in the following sense: Given a monad $(T,\eta,\mu)$ and its Kleisli category $\Kl(T)$, there is a canonical adjunction\footnote{Explicitly: The left-adjoint $L\colon \mathbf{C} \to \Kl(T)$ is given by $LX = X$ for all $\mathbf{C}$-objects $X$ and $L(f) = \eta_Y \circ f$ for all $\mathbf{C}$-arrows $f\colon X \to Y$. The right-adjoint $R\colon \Kl(T) \to \mathbf{C}$ is given by $RX = TX$ for all $\Kl(T)$-objects $X$ and $R(f)=\mu_Y \circ Tf$ for all $\Kl(T)$-arrows $f \colon X \to TY$.}
\begin{align*}
	\big(L\colon \mathbf{C} \to \Kl(T)\big) \quad  \dashv \quad \big(R\colon \Kl(T) \to \mathbf{C}\big)
\end{align*} 
with unit  $\eta'\colon \Id_\mathbf{C} \Rightarrow RL$ and counit $\epsilon\colon LR \Rightarrow  \Id_{\Kl(T)}$ giving rise to the monad, i.e. $T = RL$, $\eta=\eta'$, $\mu = R\epsilon L$. Then an endofunctor $\overline{F}$ on $\Kl(T)$ is called a \emph{lifting of $F$} if it satisfies $\overline{F}L = LF$. We will use the fact that these liftings are in one-to-one correspondence with distributive laws \cite{mulry-lifting}.

\begin{defi}[Distributive Law]
Let $(T, \eta, \mu)$ be a monad on a category $\mathbf{C}$ and $F$ be an endofunctor on $\mathbf{C}$. A natural transformation $\lambda\colon FT \Rightarrow TF$ is called a \emph{distributive law} if for all $\mathbf{C}$-objects $X$ the following diagrams commute in $\mathbf{C}$:
\[\xymatrix{
  FX 
    \ar[r]^{F\eta_X} 
    \ar[dr]_{\eta_{FX}} 	
& FTX \ar[d]^{\lambda_X} 
& 
& FT^2X	
    \ar[r]^{\lambda_{TX}} 
    \ar[d]_{F\mu_X} 	
& TFTX 
    \ar[r]^{T\lambda_X} 	
& T^2FX 
    \ar[d]^{\mu_{FX}}\\
& TFX 				
& 
& FTX		
    \ar[rr]_{\lambda_X}
&
& TFX
}\]
or equivalently $\lambda_X \circ F\eta_X = \eta_{FX}$ and $\mu_{FX} \circ T\lambda_X \circ \lambda_{TX} = \lambda_X \circ F\mu_X$.
\end{defi}

Whenever we have such a distributive law we get the lifting of a functor as defined above in the following way \cite{mulry-lifting}.

\begin{prop}[Lifting via Distributive Law]
Let $(T, \eta, \mu)$ be a monad on a category $\mathbf{C}$ and $F$ be an endofunctor on $\mathbf{C}$ with a distributive law $\lambda\colon FT \Rightarrow TF$. The distributive law induces a lifting of $F$ to an endofunctor $\overline{F}\colon \Kl(T) \to \Kl(T)$ if we define $\overline{F}X =FX$ for each object $X$ of $\Kl(T)$ and $\overline{F}(f) := \lambda_Y \circ Ff$ for each Kleisli arrow $f \colon X \to TY$. \qed
\end{prop}

\subsection{Coalgebraic Trace Semantics}

We first recall the central notions of coalgebra, coalgebra homomorphism and final coalgebra.

\begin{defi}[Coalgebra, Coalgebra-Homomorphism, Final Coalgebra]
\label{def:coalgebra}
For an endofunctor $F$ on a category $\mathbf{D}$ an $F$-coalgebra is a pair $(X, \alpha)$ where $X$ is an object and $\alpha\colon X \to FX$ is an arrow of $\mathbf{D}$. An $F$-coalgebra homomorphism between two $F$-coalgebras $(X, \alpha), (Y, \beta)$ is an arrow $\phi \colon X \to Y$ in $\mathbf{D}$ such that $\beta \circ \phi = F(\phi)\circ \alpha$. We call an $F$-coalgebra $(\Omega, \kappa)$ final if and only if for every $F$-coalgebra $(X,\alpha)$ there is a unique $F$-coalgebra-homomorphism $\phi_\alpha \colon X \to \Omega$.
\end{defi}

By choosing a suitable category and a suitable endofunctor, many (labelled) transition systems can be modelled as $F$-coalgebras. The final coalgebra -- if it exists -- can be seen as the ``universe of all possible behaviors'' and the unique map into it yields a behavioral equivalence: Two states are equivalent iff they have the same image the final coalgebra.

Whenever transition systems incorporate side-effects, these can  be ``hidden'' in a monad $T$. This leads to the following setting: the category $\mathbf{D}$ of Definition \ref{def:coalgebra} is $\Kl(T)$, i.e., the Kleisli category for the monad $T$ and a functor $\overline{F}\colon \Kl(T)\to \Kl(T)$ is obtained by suitably lifting a functor $F$ of the underlying category (such that $\overline{F}X = FX$ on objects, see above). Then coalgebras are defined as arrows $\alpha\colon X\to\overline{F}X$ in the Kleisli category, which can be regarded as arrows $X\to TFX$ in the base category. As indicated in the introduction, the monad can be seen as describing implicit branching (side effects), whereas $F$ describes the explicit branching structure.

In this setup the final coalgebra in the Kleisli category often yields a notion of trace semantics \cite{hasuo,Sokolova20115095}. The side effects specified via the  monad are not part of the final coalgebra, but are contained in the unique map into the final coalgebra (which is again a Kleisli arrow).

In our case $T$ is either the sub-probability or the probability monad on $\mathbf{Meas}$ (which will be defined later), whereas $F$ is defined as $F = \A \times \Id_\Meas + \one$ or $F = \A \times \Id_\Meas$ for a given finite alphabet $\A$. That is, the monad $T$ describes probabilistic branching, whereas the endofunctor $F$ specifies (explicitly observable) labels and possibly termination.

\subsection{Borel-Sigma-Algebras and the Lebesgue Integral}
Before we can define the probability and the sub-probability monad, we give a crash course in integration loosely based on \cite{ash,Els07}. For that purpose let us fix a measurable space $X$ and a measure $\mu$ on $X$. We want to integrate numerical functions $f\colon X \to \ExtR$ and in order to do that we need a suitable $\sigma$-algebra on $\ExtR$ to define measurability of such functions.

Recall that a topological space is a tuple $(Y, \mathcal{T})$, where $Y$ is a set and $\mathcal{T} \subseteq \powerset{Y}$ is a set containing the empty set, the set $Y$ itself and is closed under arbitrary unions and finite intersections. The set $\mathcal{T}$ is called the \emph{topology} of $Y$ and its elements are called \emph{open sets}. The \emph{Borel $\sigma$-algebra} on $Y$, denoted $\Borel(Y)$, is the $\sigma$-algebra generated by the open sets $\mathcal{T}$ of the topology, i.e. $\Borel(Y) = \sigalg[Y]{\mathcal{T}}$. Thus the Borel $\sigma$-algebra provides a connection of topological aspects and measurability. For the set of real numbers, it can be shown (\cite[I.4.3 Satz]{Els07}) that the Borel $\sigma$-algebra $\Borel(\R)$ is generated by the semiring of all left-open intervals
\begin{align*}
	\Borel(\R) = \sigalg[\R]{\set{\,(a,b]\mid a,b \in \R, a \leq b}}. 
\end{align*}
With this definition at hand, we now equip the set $\ExtR$ of extended reals with its Borel \linebreak $\sigma$-algebra which can be defined as
\begin{align*}
	\Borel(\ExtR) = \sigalg[\ExtR]{\set{B \cup E \mid B \in \Borel(\R), E \subseteq \set{-\infty, \infty}}}.
\end{align*}
A function $f\colon X \to \ExtR$ is called \emph{(Borel-)measurable} if it is measurable with with respect to this Borel $\sigma$-algebra. Given two Borel-measurable functions $f,g \colon Y \to \ExtR$ and real numbers $\alpha, \beta$ also $\alpha f+\beta g$ is Borel-measurable \cite[III.4.7]{Els07} and thus are all finite linear combinations of Borel-measurable functions. Moreover, if $(f_n)_{n \in \N}$ is a sequence of Borel-measurable functions $f_n\colon X \to \ExtR$ converging pointwise to a function $f\colon X \to \ExtR$, then also $f$ is Borel-measurable \cite[1.5.4]{ash}. In the remainder of this section we will just consider Borel-measurable functions. 

We call $f$ \emph{simple} iff it attains only finitely many values, say $f(X) = \set{\alpha_1, \hdots, \alpha_N}$. The integral of such a simple function $f$ is then defined to be the $\mu$-weighted sum of the $\alpha_n$, formally $\Int{f}[\mu] = \sum_{n=1}^{N}\alpha_n\mu(S_n)$ where $S_n = f^{-1}(\alpha_n) \in \Sigma_X$. Whenever $f$ is non-negative we can approximate it from below using non-negative simple functions. In this case we define the integral to be 
\[\Int{f}[\mu] := \sup\set{\Int{s}[\mu]\mid s \mbox{ non-negative and simple s.t. } 0 \leq s \leq f}.\]
 For arbitrary Borel-measurable $f$ we decompose it into its positive part $f^+ := \max\set{f,0}$ and negative part $f^- := \max\set{-f, 0}$ which are both non-negative and Borel-measurable. We note that $f = f^+ - f^{-}$ and consequently we define the integral of $f$ to be the difference $\Int{f}[\mu] := \Int{f^+}[\mu]  - \Int{f^-}[\mu]$ if not both integrals on the right hand side are $+\infty$. In the latter case we say that the integral does not exist. Whenever it exists and is finite we call $f$ a \emph{$\mu$-integrable function} or simply an \emph{integrable function} if the measure $\mu$ is obvious from the context.

For every measurable set $S \in \Sigma_X$ its characteristic function $\chi_S \colon X \to \R$, which is \linebreak $1$ if $x \in S$ and $0$ otherwise, is $\mu$-integrable and for $\mu$-integrable $f$ the product $\chi_S \cdot f$ is also $\mu$-integrable and we write 
\begin{align*}	
	\Int[S]{f}[\mu] := \Int{\chi_S \cdot f}[\mu] \,.
\end{align*}
 Instead of $\Int[S]{f}[\mu]$ we will sometimes write $\Int[S]{f(x)}[\mu(x)]$ or $\Int[{x \in S}]{f(x)}[\mu(x)]$  which is useful if we have functions with more than one argument or multiple integrals. Note that this does not imply that singleton sets are measurable. 

Some useful properties of the integral are that it is \emph{linear}, i.e. for $\mu$-integrable functions $f,g\colon X \to \ExtR$ and real numbers $\alpha, \beta$ we have 
\begin{align*}
	\Int{\alpha f + \beta g}[\mu] = \alpha \Int{f}[\mu] + \beta\Int{g}[\mu]
\end{align*}
and the integral is \emph{monotone}, i.e. $f \leq g$ implies $\Int{f}[\mu]  \leq \Int{g}[\mu] $.  We will state one result explicitly which we will use later in our proofs. This result and its proof can be found e.g. in \cite[Theorem 1.6.12]{ash}.

\begin{prop}[Image Measure]
Let $X, Y$ be measurable spaces, $\mu$ be a measure on $X$, $f\colon Y \to \ExtR$ be a Borel-measurable function and $g\colon X \to Y$ be a measurable function. Then $\mu \circ g^{-1}$ is a measure\footnote{This notation is a bit lax, if we wanted to be really precise we would have to write $\mu \circ \left(g^{-1}|_{\Sigma_Y}\right)$.} on $Y$, the so-called \emph{image-measure} and $f$ is $(\mu \circ g^{-1})$-integrable iff $f \circ g$ is $\mu$-integrable and in this case we have $\Int[S]{f}[(\mu \circ g^{-1})] = \Int[{g^{-1}(S)}]{f \circ g}[\mu]$ for all $S \in \Sigma_Y$.\qed
\end{prop}

\subsection{The Probability and the Sub-Probability Monad}
We will now introduce the probability monad (Giry monad) and the sub-probability monad as e.g. presented in \cite{Gir82} and \cite{Pan09}. First, we take a look at the endofunctors of these monads.

\begin{defi}[The Sub-Probability and the Probability Functor]
The \emph{sub-probability-functor} $\mathbb{S} \colon \Meas \to \Meas$ maps a measurable space $(X, \Sigma_X)$ to the measurable space  $\big(\mathbb{S}(X), \Sigma_{\mathbb{S}(X)}\big)$ where
 $\mathbb{S}(X)$ is the set of all sub-probability measures on $\Sigma_X$ and $\Sigma_{\mathbb{S}(X)}$ is the smallest  $\sigma$-algebra such that for all $S \in \Sigma_X$ the \emph{evaluation maps}: 
\begin{align}
	\quad p_S\colon \mathbb{S}(X) \to [0,1],\quad p_S(P) = P(S) \label{eq:evaluation_map}
\end{align}
are Borel-measurable. For any measurable function $f \colon X \to Y$ between measurable spaces $(X, \Sigma_X)$, $(Y, \Sigma_Y)$ the arrow $\mathbb{S}(f)$ maps a probability measure $P$ to its image measure:
\begin{align}
	\mathbb{S}(f) \colon \mathbb{S}(X) \to \mathbb{S}(Y), \quad \mathbb{S}(f)(P) := P \circ f^{-1}.
\end{align}
If we take full probabilities instead of sub-probabilities we get another endofunctor, the probability functor $\mathbb{P}$, analogously.
\end{defi}

Both the sub-probability functor $\mathbb{S}$ and the probability functor $\mathbb{P}$ are functors of monads with the following unit and multiplication natural transformations.

\begin{defi}[Unit and Multiplication]
Let $T$ be either the sub-probability functor $\mathbb{S}$ or the probability functor $\mathbb{P}$. We obtain two natural transformations $\eta \colon \Id_\Meas \Rightarrow T$ and $\mu \colon T^2\Rightarrow T$ by defining for every measurable space $(X,\Sigma_X)$:
\begin{align}
	\eta_X \colon X \to TX,\ & \quad \eta_X(x) = \delta_x^X \label{eq:giry_unit}\\
	\mu_X \colon T^2X \to TX,\ & \quad \mu_X(P)(S) := \Int{p_S}[P] \quad \text{for } S \in \Sigma_X\label{eq:giry_mult}
\end{align}
 where $\delta_x^X\colon \Sigma_X \to [0,1]$ is the Dirac measure and $p_S$ is the evaluation map \eqref{eq:evaluation_map} from above.
\end{defi}

If we combine all the ingredients we obtain the following result which also guarantees the soundness of the previous definitions.

\begin{prop}[\cite{Gir82,Pan09}]
 $(\mathbb{S}, \eta, \mu)$ and $(\mathbb{P}, \eta, \mu)$ are monads on $\Meas$.\qed
\end{prop}

\subsection{A Category of Stochastic Relations}
The Kleisli category of the sub-probability monad $(\mathbb{S}, \eta, \mu)$ is sometimes called \emph{category of stochastic relations \cite{Pan09}} and denoted by $\mathbf{SRel}$. Let us briefly analyze the arrows of this category: Given two measurable spaces $(X,\Sigma_X)$, $(Y, \Sigma_Y)$ a Kleisli arrow $h \colon X \to \mathbb{S}Y$ maps each $x \in X$ to a sub-probability measure $h(x) \colon \Sigma_Y \to [0,1]$. By uncurrying we can regard $h$ as a function $h\colon X \times \Sigma_Y\to [0,1]$. Certainly for each $x \in X$ the function $S \mapsto h(x,S)$ is a (sub-)probability measure and one can show that for each $S \in \Sigma_Y$ the function $x \mapsto h(x,S)$ is Borel-measurable. Any function $h \colon X \times \Sigma_Y\to [0,1]$ with these properties is called a \emph{Markov kernel} or a \emph{stochastic kernel} and it is known \cite[Proposition 2.7]{Dob07b} that these Markov kernels correspond exactly to the Kleisli arrows $h \colon X \to \mathbb{S}Y$. 

We will later need the following, simple result about Borel-measurable functions and Markov kernels:

\begin{lem}
	\label{lem:measMarkovKernel}
	Let $(X, \Sigma_X)$ and $(Y, \Sigma_Y)$ be measurable spaces, $g \colon Y \to [0,1]$ be a Borel-measurable function and $h\colon X \times \Sigma_Y \to [0,1]$ be a Markov kernel. Then the function\linebreak $f\colon X \to [0,1]$, $f(x) := \Int[y \in Y]{g(y)}[h(x,y)]$ is Borel-measurable.
\end{lem}

\proof
If $g$ is a simple and Borel-measurable function, say $g(Y) = \set{\alpha_1,..., \alpha_N}$, then $f(x) = \sum_{n=1}^N \alpha_n h(x,A_n)$ where $A_n = g^{-1}(\set{\alpha_n})$ and hence $f$ is Borel-measurable as a linear combination of Borel-measurable functions. If $g$ is an arbitrary, Borel-measurable function we approximate it from below with simple functions $s_i$, $i \in \N$ and define $f_i\colon X \to [0,1]$ with $f_i(x) = \Int[y \in Y]{s_i(y)}[h(x,y)]$. Then by the monotone convergence theorem (\cite[1.6.2]{ash}) we have $f(x) = \Int[y \in Y]{\lim_{i \to \infty}s_i(y)}[h(x,y)] = \lim_{i \to \infty}f_i(x)$. As shown before, each of the $f_i$ is Borel-measurable and thus also the function $f$ is Borel-measurable as pointwise limit of Borel-measurable functions.
\qed

\section{Main Results}
\subsection{Continuous Probabilistic Transition Systems}
There is a big variety of probabilistic transition systems \cite{Sokolova20115095,glabbeek}. We will deal with four slightly different versions of so-called \emph{generative} PTS. The underlying intuition is that, according to a sub-probability measure, an action from the alphabet $\mathcal{A}$ and a set of possible successor states are chosen. We distinguish between probabilistic branching according to sub-probability and probability measures and furthermore we treat systems without and with termination. 

\begin{defi}[Probabilistic Transition System]
A \emph{probabilistic transition system}, short \emph{PTS}, is a tuple $(\A, X, \alpha)$ where $\A$ is a finite alphabet (endowed with $\powerset{\A}$ as $\sigma$-algebra), $X$ is the \emph{state space}, an arbitrary measurable space with $\sigma$-algebra $\Sigma_X$ and  $\alpha$ is the \emph{transition function} which has one of the following forms and determines the type\footnote{The reason for choosing these symbols as type-identifiers will be revealed later in this paper.} of the PTS.
\begin{center}\begin{tabular}{p{5cm}|c}
	\hline
	Transition Function $\alpha$ 						& Type $\diamond$ of the PTS\\
	\hline
	$\alpha\colon X \to \mathbb{S}(\A \times X)$ 			& $0$\\
	$\alpha\colon X \to \mathbb{S}(\A \times X + \one)$ 		& $*$\\
	$\alpha\colon X \to \mathbb{P}(\A \times X)$ 			& $\omega$\\
	$\alpha\colon X \to \mathbb{P}(\A \times X + \one)$ 		& $\infty$\\
	\hline
\end{tabular}\end{center}
For every symbol $a \in \A$ we define a Markov kernel  $\mathbf{P}_{a}\colon X \times \Sigma_X \to [0,1]$ where 
\begin{align}
	 \P{a}{x}{S} := \alpha(x)(\set{a} \times S)\,.
\end{align}
Intuitively, $\P{a}{x}{S}$ is the probability of making an $a$-transition from the state $x \in X$ to any state $y \in S$. Whenever $X$ is a countable set and $\Sigma_X = \powerset{X}$ we call the PTS \emph{discrete}. The unique state $\checkmark \in \one$ -- whenever it is present -- denotes termination of the system. 
\end{defi}

We will now take a look at a small example $\infty$-PTS before we continue with our theory. 

\begin{exa}[Discrete PTS with Finite and Infinite Traces]
\label{ex:pts}
Let $\A = \set{a,b}$, $X = \set{0,1,2}$, $\Sigma_X = \powerset{X}$ and $\alpha \colon X \to \mathbb{P}(\A \times X + \one)$ such that we obtain the following system.
\begin{center}
\begin{tikzpicture}[node distance=1.4 and 2.8, on grid, shorten >=1pt,
    >=stealth', semithick]
\node[state, inner sep=2pt, minimum size=20pt,draw](q0) {$0$};
\node[state, inner sep=2pt, minimum size=20pt,draw, right=of q0] (q1) {$1$}; 
\node[state, inner sep=2pt, minimum size=20pt,draw, below=of q1] (q2) {$2$};
\node[state, inner sep=2pt, minimum size=20pt,draw, right=of q1, accepting] (q3) {$\checkmark$};
\draw[->] (q0) edge[loop left] node[left] {$b,1$} (q0);
\draw[->] (q1) edge node[above] {$b, 1/3$} (q0);
\draw[->] (q1) edge node[left] {$a, 1/3$} (q2);
\draw[->] (q1) edge node[above] {$1/3$} (q3);
\draw[->] (q2) edge[loop left] node[left] {$a, 2/3$} (q2);
\draw[->] (q2) edge node[below] {$1/3$} (q3);
\end{tikzpicture}
\end{center}

\noindent As stated in the definition, $\checkmark$ is the unique final state. It has only incoming transitions bearing probabilities and no labels. The intuitive interpretation of these transitions can be stated as follows: ``From state $1$ the system terminates immediately with probability $1/3$''.   
\end{exa}

\subsection{Towards Measurable Sets of Words: Cones and Semirings}
\label{sec:cones}
In order to define a trace measure on these probabilistic transition systems we need suitable $\sigma$-algebras on the sets of words. While the set of all finite words, $\Astar$, is rather simple -- we will take $\powerset{\Astar}$ as $\sigma$-algebra -- the set of all infinite words, $\Aomega$, and also the set of all finite and infinite words, $\Ainfty$, needs some consideration. For a word $u \in \Astar$ we call the set of all infinite words that have $u$ as a prefix the \emph{$\omega$-cone} of $u$, denoted by $\cone{\omega}{u}$, and similarly we call the set of all finite and infinite words having $u$ as a prefix the \emph{$\infty$-cone} \cite[p.~23]{Pan09} of $u$ and denote it with $\cone{\infty}{u}$. Sometimes, e.g. in \cite{baierkatoen2008}, these sets are also-called ``cylinder sets''. 

A cone can be visualized in the following way: For a given alphabet $\A \not = \emptyset$ we consider the undirected, rooted and labelled tree given by $\mathcal{T} := (V, E,\epsilon, l)$ with infinitely many vertices $V := \Astar$, edges $E := \set{ \set{u, ua}\mid u \in \mathcal{A}^*, a \in \mathcal{A}}$, root $\epsilon \in \mathcal{A}^*$ and edge-labeling function $l \colon E \to \mathcal{A}, \set{u, ua} \mapsto a$. For $\mathcal{A} = \set{a,b,c}$ the first three levels of the tree can be depicted as follows:
\[\begin{xy}\xymatrix{
		&	&&& \ar@{-}[dlll]_a\epsilon\ar@{-}[d]_b\ar@{-}[drrr]^c\\
		& \ar@{-}[dl]_a a \ar@{-}[d]_b\ar@{-}[dr]^c &&& \ar@{-}[dl]_a b \ar@{-}[d]_b\ar@{-}[dr]^c &&& \ar@{-}[dl]_a c \ar@{-}[d]_b\ar@{-}[dr]^c\\ 
	aa	& ab & ac & ba & bb & bc & ca & cb & cc 
}\end{xy}
\]
Given a finite word $u \in \mathcal{A}^*$, the $\omega$-cone of $u$ is represented by the set of all infinite paths\footnote{Within this paper a path of an undirected graph $(V,E)$ is always considered to be \emph{simple}, i.e. any two vertices in a path are different.} that begin in $\epsilon$ and contain the vertex $u$ and the $\infty$-cone of $u$ is represented by the set of all finite and infinite paths that begin in $\epsilon$ and contain the vertex $u$ (and thus necessarily have a length which is greater or equal to the length of $u$). 

\begin{defi}[Cones]
Let $\A$ be a finite alphabet and let $\sqsubseteq \, \subset \Astar \times \Ainfty$ denote the usual prefix relation on words. For $u \in \Astar$ we define its  $\omega$-\emph{cone} to be the set $\cone{\omega}{u}:=\set{v \in \Aomega\mid u \sqsubseteq v }$ and analogously we define $\cone{\infty}{u}:=\set{v \in \Ainfty\mid u \sqsubseteq v }$, the $\infty$-\emph{cone} of $u$. 
\end{defi}

With this definition at hand, we can now define the semirings we will use to generate $\sigma$-algebras on $\emptyset$, $\Astar$, $\Aomega$ and $\Ainfty$.

\begin{defi}[Semirings of Sets of Words]
\label{def:semirings_of_words}
Let $\A$ be a finite alphabet. We define 
\begin{align*}
	\S_0 &:= \set{\emptyset} \subset \powerset{\emptyset},\\
	\S_* &:= \set{\emptyset}\cup \set{\set{u}\mid u \in \Astar} \subset \powerset{\A^*},\\
	\S_\omega &:= \set{\emptyset}\cup \set{\cone{\omega}{u}\mid u \in \Astar} \subset \powerset{\A^\omega},\\
	\Sinfty &:= \set{\emptyset}\cup \set{\set{u}\mid u \in \Astar} \cup \set{\cone{\infty}{u}\mid u \in \Astar}\subset \powerset{\Ainfty}.
\end{align*}
\end{defi}
\noindent For the next proposition the fact that $\A$ is a finite alphabet is crucial.
\begin{prop}
\label{prop:semirings_of_words}
The sets $\S_0$, $\Sstar$, $\Somega$ and $\Sinfty$ are covering semirings of sets.
\end{prop}
\proof
For $\S_0 = \set{\emptyset}$ nothing has to be shown. Obviously we have $\emptyset \in \Sstar$ and for elements $\set{u}, \set{v} \in \Sstar$ we remark that $\set{u} \cap \set{v}$ is either $\set{u}$ iff $u = v$ or $\emptyset$ else. Moreover, $\set{u} \setminus \set{v}$ is either $\emptyset$ iff $u=v$ or $\set{u}$ else. We proceed with the proof for $\Sinfty$, the proof for $\Somega$ can be carried out almost analogously (in fact, it is simpler). By definition we have $\emptyset \in \Sinfty$. An intersection $\cone{\infty}{u} \,\cap \cone{\infty}{v}$ is non-empty iff either $u \sqsubseteq v$ or $v \sqsubseteq u$ and is then equal to $\cone{\infty}{v}$ or to $\cone{\infty}{u}$ and thus an element of $\mathcal{S}_\infty$. Similarly an intersection $\cone{\infty}{u} \cap \set{v}$ is non-empty iff $u \sqsubseteq v$ and is then equal to $\set{v} \in \Sinfty$. As before we have $\set{u} \cap \set{v} = \set{u}$ for $u=v$ and $\set{u} \cap \set{v} = \emptyset$ else.  For the set difference $\cone{\infty}{u} \,\setminus \cone{\infty}{v}$ we denote that this is either $\emptyset$ (iff $v \sqsubseteq u$) or $\cone{\infty}{u}$ (iff  $v \not \sqsubseteq u$ and $u \not \sqsubseteq v$) or otherwise ($u \sqsubseteq v$) the following union\footnote{For $n \in \N$ we define $\A^{<n} := \set{u \in \A \mid |u| < n}$.} of finitely many disjoint sets in $\Sinfty$:
\begin{align*}
	\cone{\infty}{u} \,\setminus \cone{\infty}{v} = \left(\bigcup\limits_{v' \in \A^{|v|}\setminus\set{v}, u \sqsubseteq v'} \! \cone{\infty}{v'}\right) \cup \left(\bigcup\limits_{v' \in \A^{< |v|},~u \sqsubseteq v'}\set{v'}\right).
\end{align*}
As before we get $\set{u}\setminus\set{v}=\emptyset$ iff $u=v$ and $\set{u}\setminus\set{v} = \set{u}$ else. For $\set{u} \setminus \cone{\infty}{v}$ we observe that this is either $\set{u}$ iff $v \not \sqsubseteq u$ or $\emptyset$ else. Finally, $\cone{\infty}{u} \,\setminus \set{v}$ is either $\cone{\infty}{u}$ (iff $u \not \sqsubseteq v$) or ($u \sqsubseteq v$) the following union of finitely many disjoint sets in $\Sinfty$:
\begin{align*}
	\cone{\infty}{u} \,\setminus \set{v} = \left(\bigcup\limits_{v' \in \A^{|v|}\setminus\set{v}, u \sqsubseteq v'} \! \cone{\infty}{v'}\right) \cup \left(\bigcup\limits_{v' \in \A^{< |v|},~u \sqsubseteq v'}\set{v'}\right) \cup \left(\bigcup\limits_{a \in \A} \! \cone{\infty}{va}\right)
\end{align*}
which completes the proof that the given sets are semirings. The countable (and even disjoint) covers are: $\emptyset = \emptyset$, $\Astar = \cup_{u \in \Astar}\set{a}$, $\Aomega =\, \cone{\omega}{\epsilon}$ and $\Ainfty =\, \cone{\infty}{\epsilon}$.
\qed

We remark that many interesting sets will be measurable in the $\sigma$-algebra generated by these cones. The singleton-set $\set{u}$ will be measurable for every $u \in \Aomega$ because $\set{u} = \bigcap_{v \sqsubseteq u}\cone{\omega}{v} = \bigcap_{v \sqsubseteq u}\cone{\infty}{v}$ which are countable intersections, and (for $\infty$-cones only) the set $\Astar = \cup_{u \in \Astar}\set{u}$ and consequently also the set $\Aomega = \Ainfty \setminus \Astar$ will be measurable. The latter will be useful to check to what ``extent'' a state of a $\infty$-PTS accepts finite or infinite behavior.

\subsection{Measurable Sets of Words}
Let us now take a closer look at the $\sigma$-algebras generated by the semirings which we defined in the last section. We obviously obtain the trivial $\sigma$-algebra $\sigalg[\emptyset]{\S_0} = \set{\emptyset}$. Since $\A$ is finite, $\Astar$ is countable and we can easily conclude $\sigalg[\Astar]{\Sstar} = \powerset{\Astar}$. The other two cases need a more thorough treatment. For the remainder of this section let thus $\diamond \in \set{\omega,\infty}$. We will use the concepts of transfinite induction (cf. e.g. to \cite{Dud89} for an introduction) to extend the semi-ring $\S_\diamond$ to the $\sigma$-algebra it generates. A similar construction is well-known and presented e.g. in \cite{Els07}. Usually this explicit construction is not needed but for our proofs it will turn out to be useful. 
\begin{defi}
	For any set $X$ and $\mathcal{G} \subseteq \powerset{X}$ let $\Unions{\mathcal{G}}$ and $\Intersections{\mathcal{G}}$ be the closure of $\mathcal{G}$ under countable unions and intersections. We define $\Rdiamond(0) :=\set{\cup_{n=1}^N S_n \mid N \in \N, S_n \in \S_\diamond \text{ disjoint}}$, $\Rdiamond(\alpha+1) := \Unions{\Intersections{\Rdiamond(\alpha)}}$ for every ordinal $\alpha$ and $\Rdiamond(\gamma) := \cup_{\alpha < \gamma} \Rdiamond(\alpha)$ for every limit ordinal $\gamma$. 
\end{defi}
Obviously we have $\Rdiamond(\alpha) \subseteq \Rdiamond(\beta)$ for all ordinals $\alpha < \beta$. Since $\S_\diamond$ is a semiring of sets, is easy to see that $\Rdiamond(0)$ is an \emph{algebra}, i.e. it contains the base set $\A^\diamond$, is closed under complement and binary (and hence all finite) unions and intersections.  

\begin{lem}
	\label{lem:complementsLimit}
	$A \in \Rdiamond(\gamma) \implies \compl{A} \in \Rdiamond(\gamma)$ for every limit ordinal $\gamma$.
\end{lem}
\proof
	We will show that $A \in \Rdiamond(\alpha) \implies \compl{A} \in \Intersections{\Rdiamond(\alpha)}$ for every ordinal $\alpha$. This is true for the algebra $\Rdiamond(0)$. Now let $\alpha$ be an ordinal satisfying the implication and let $A \in \Rdiamond(\alpha+1)$. Then $A = \cup_{m=1}^\infty \cap_{n=1}^\infty A_{m,n}$ with $A_{m,n} \in \Rdiamond(\alpha)$ and by deMorgan's rules $\compl{A} = \cap_{m=1}^\infty \cup_{n=1}^\infty \compl{A_{m,n}}$ where by hypothesis $\compl{A_{m,n}} \in \Intersections{\Rdiamond(\alpha)}$, thus $\cup_{n=1}^\infty \compl{A_{m,n}} \in \Unions{\Intersections{\Rdiamond(\alpha)}} = \Rdiamond(\alpha+1)$ and therefore $\compl{A} \in \Intersections{\Rdiamond(\alpha+1)}$. Finally, let $\gamma$ be a limit ordinal and suppose the implication holds for all ordinals $\alpha < \gamma$. For any $B\in \Rdiamond(\gamma)$ there is a  $\beta < \gamma$ such that $B \in \Rdiamond(\beta)$. Hence we have $\overline{B} \in \Intersections{\Rdiamond(\beta)} \subseteq \Intersections{\Rdiamond(\gamma)} \subseteq \Rdiamond(\gamma)$.
\qed

\begin{lem}
\label{lem:finite_union_intersection}
$A, B \in \Rdiamond(\alpha) \implies A\cup B, A \cap B \in \Rdiamond(\alpha)$ for every ordinal $\alpha$.
\end{lem}
\proof
This is true for the algebra $\Rdiamond(0)$. Let $\alpha$ be an ordinal satisfying the implication and $A, B \in \Rdiamond(\alpha+1)$, then $A = \cup_{k=1}^\infty \cap_{l=1}^\infty A_{k,l}$ and $B = \cup_{m=1}^\infty \cap_{n=1}^\infty B_{m,n}$ with $A_{k,l}, B_{m,n} \in \Rdiamond(\alpha)$. Obviously $A \cup B = \cup_{k,m=1}^\infty \cap_{l,n=1}^\infty  ( A_{k,l} \cup B_{m,n})$ and $A \cap B = \cup_{k,m=1}^\infty \cap_{l,n=1}^\infty ( A_{k,l} \cap B_{m,n})$ where by hypothesis $A_{k,l} \cup B_{m,n}, A_{k,l} \cap B_{m,n} \in \Rdiamond(\alpha)$. Let $\gamma$ be a limit ordinal and suppose the statement is true for all $\alpha < \gamma$ and let $A, B \in \Rdiamond(\gamma)$. There must be ordinals $\alpha, \beta < \gamma$ such that $A \in \Rdiamond(\alpha)$ and $B \in \Rdiamond(\beta)$. Assume wlog $\alpha \leq \beta$ then $A \in \Rdiamond(\beta)$, hence $A \cup B, A \cap B \in \Rdiamond(\beta) \subseteq \Rdiamond(\gamma)$ which completes the proof. 
\qed

\begin{lem}
	\label{lem:IntersectionIsClosedUnderFiniteUnion}
	$A, B \in \Intersections{\Rdiamond(\alpha)} \implies A\cup B \in \Intersections{\Rdiamond(\alpha)}$ for every ordinal $\alpha$.
\end{lem}
\proof
	Let $A, B \in \Intersections{\Rdiamond(\alpha)}$ then $A:= \cap_{m=1}^\infty A_m$ and $B:= \cap_{n=1}^\infty B_n$ with $A_m, B_n \in \Rdiamond(\alpha)$. Then $A \cup B = \cap_{m,n=1}^\infty (A_m \cup B_n)$ where $A_m \cup B_n \in \Rdiamond(\alpha)$ by Lemma \ref{lem:finite_union_intersection} and thus \linebreak $A \cup B \in \Intersections{\Rdiamond(\alpha)}$.
\qed

\begin{prop}
	\label{prop:TransFiniteSigAlg}
	$\sigalg[\A^\diamond]{\Rdiamond(0)} = \Rdiamond(\omega_1)$ where $\omega_1$ is the smallest uncountable limit ordinal.
\end{prop}
\proof[Proof (adapted from \cite{Els07}).] We first show $\Rdiamond(\omega_1) \subseteq \sigalg[X]{\Rdiamond(0)}$. We know that \linebreak $\Rdiamond(0) \subseteq \sigalg[X]{\Rdiamond(0)}$. For an ordinal $\alpha$ with $\Rdiamond(\alpha) \subseteq \sigalg[X]{\Rdiamond(0)}$ let $A \in \Rdiamond(\alpha+1)$. Then $A = \cup_{m=1}^\infty\cap_{n=1}^\infty A_{m,n}$ with $A_{m,n} \in \Rdiamond(\alpha)$ yielding $A \in  \sigalg[X]{\Rdiamond(0)}$. If $\gamma$ is a limit ordinal with $\Rdiamond(\alpha) \subseteq \sigalg[X]{\Rdiamond(0)}$ for all ordinals $\alpha < \gamma$ then for any $A \in \Rdiamond(\gamma)$ there must be an ordinal $\alpha < \gamma$ such that $A \in \Rdiamond(\alpha)$ and hence $A \in \sigalg[X]{\Rdiamond(0)}$. In order to show $\Rdiamond(\omega_1) \supseteq \sigalg[X]{\Rdiamond(0)}$ it suffices to show that $\Rdiamond(\omega_1)$ is a $\sigma$-algebra. We have $X \in R(0) \subseteq R(\omega_1)$ and Lemma \ref{lem:complementsLimit} yields closure under complements. Let $A_n \in \Rdiamond(\omega_1)$ for $n \in \N$. Then for each $A_n$ we have an $\alpha_n$ such that $A_n \in \Rdiamond(\alpha_n)$. Since $\omega_1$ is the first uncountable ordinal, we must find an $\alpha < \omega_1$ such that $\alpha_n < \alpha$ for all $n \in \N$. Hence we have $A_n \in \Rdiamond(\alpha)$ for all $n \in \N$. Thus $\cup_{n=1}^\infty A_n \in \Rdiamond(\alpha+1) \subseteq \Rdiamond(\omega_1)$.
\qed

\subsection{The Trace Measure}
We will now define the trace measure which can be understood as the behavior of a state: it measures the probability of accepting a set of words.

\begin{defi}[The Trace Measure]
\label{def:trace_premeasure}
Let $(\A, X, \alpha)$ be a $\diamond$-PTS. For every state $x \in X$ we define the trace (sub-)probability measure $\tr(x) \colon \sigalg[\A^\diamond]{\mathcal{S}_\diamond} \to [0,1]$ as follows: In all four cases we require $\tr(x)(\emptyset) = 0$. For $\diamond \in \set{*, \infty}$ we define 
\begin{align}
	\tr(x)(\set{\epsilon}) = \alpha(x)(\one) \label{eq:trace_emptyword}
\end{align} and
\begin{align}
	\tr(x)\big(\set{au}\big) := \Int[{x' \in X}]{\tr(x')(\set{u})}[\P{a}{x}{x'}]  \label{eq:trace_main_equation}
\end{align} 
for all $a \in A$ and all $u \in \Astar$. For $\diamond \in \set{\omega, \infty}$ we define 
\begin{align}
	\tr(x)(\cone{\diamond}{\epsilon}) = 1\label{eq:trace_wholespace}
\end{align} and
\begin{align}
	\tr(x)\big(\cone{\diamond}{au}\big) := \Int[{x' \in X}]{\tr(x')(\cone{\diamond}{u})}[\P{a}{x}{x'}] \label{eq:trace_main_equation2}
\end{align} 
for all $a \in A$ and all $u \in \Astar$. 
\end{defi}

We need to verify that everything is well-defined and sound. In the next proposition we explicitly state what has to be shown.

\begin{prop}
\label{prop:trace_premeasure}
For all four types $\diamond \in \set{0,*,\omega, \infty}$ of PTS the equations in Definition~\ref{def:trace_premeasure} yield a $\sigma$-finite pre-measure  $\tr(x)\colon \S_\diamond \to [0,1]$ for every $x \in X$. Moreover, the unique extension of this pre-measure is a (sub-)probability measure.
\end{prop}

Before we prove this proposition, let us try to get a more intuitive understanding of Definition~\ref{def:trace_premeasure} and especially equation \eqref{eq:trace_main_equation}. First we check how the above definition reduces when we consider discrete systems.  

\begin{rem}
Let $(\A, X, \alpha)$ be a discrete\footnote{If $Z$ is a countable set and $\mu\colon \powerset{Z} \to [0,1]$ is a measure, we write $\mu(z)$ for $\mu(\set{z})$.} $*$-PTS, i.e. $X$ is a countable set with \linebreak $\sigma$-algebra $\powerset{X}$ and the transition probability function is $\alpha \colon X \to \mathbb{S}(\A \times X + \one)$. Then  $\tr(x)(\epsilon) := \alpha(x)(\checkmark)$ and \eqref{eq:trace_main_equation} is equivalent to
\begin{align}
	 \quad \tr(x)(au) := \sum_{x' \in X} \tr(x')(u) \cdot \P{a}{x}{x'}
\end{align}
for all $a \in \A$ and all $u \in \Astar$ which in turn is equivalent to the discrete ``trace distribution'' presented in \cite{Hasuo06generictrace} for the sub\--dis\-tri\-bu\-tion monad $\mathcal{D}$ on $\SeT$. 
\end{rem}

Having seen this coincidence with known results, we proceed to calculate the trace measure for our example (Example \ref{ex:pts}) which we can only do in our more general setting because this $\infty$-PTS is a discrete probabilistic transition system which exhibits both finite and infinite behavior. 

\begin{exa}[Example \ref{ex:pts} continued.]
We calculate the trace measures for the $\infty$-PTS from Example \ref{ex:pts}. We have $\tr(0) = \delta_{b^\omega}^\Ainfty$ because 
\begin{align*}
\tr(0)(\set{b^\omega}) &= \tr(0)\left(\bigcap_{k=0}^{\infty}\cone{\infty}{b^k}\right)=\tr(0)\left(\Ainfty \setminus \bigcup_{k=0}^{\infty}\left(\Ainfty \setminus \cone{\infty}{b^k}\right)\right) \\
	&= \tr(0)\left(\Ainfty\right) - \tr(0)\left(\bigcup_{k=0}^{\infty}\left(\Ainfty \setminus \cone{\infty}{b^k}\right)\right) \geq 1 - \sum_{k=0}^{\infty}\tr(0)\left(\Ainfty \setminus \cone{\infty}{b^k}\right)\\
	&= 1 - \sum_{k=0}^{\infty} \left(1- \tr(0)\left(\cone{\infty}{b^k}\right)\right) = 1-\sum_{k=0}^{\infty}(1-1) = 1
\end{align*} 
Thus we have $\tr(0)(\Astar) = \tr(0)\left(\cup_{u \in \Astar}\set{u}\right) = 0$ and $\tr(0)(\Aomega) = 1$. By induction we can show that $\tr(2)(\set{a^k}) = (1/3) \cdot (2/3)^k$ and thus $\tr(2)(\Astar) = 1$ because
\begin{align*}
	1 \geq \tr(2)(\Astar) = \tr(2)\left(\bigcup_{u \in \Astar}^\infty \set{u}\right)\geq \tr(2)\left(\bigcup_{k=0}^\infty \set{a^k}\right)= \frac{1}{3}\cdot\sum_{k=0}^\infty\left(\frac{2}{3}\right)^k = 1
\end{align*} 
and hence $\tr(2)(\Aomega) = 0$. Furthermore we calculate $\tr(1)(\set{b^\omega})= 1/3$, $\tr(1)(\cone{\infty}{a}) = 1/3$ and $\tr(1)(\set{\epsilon}) = 1/3$ yielding $\tr(1)(\Astar) = 2/3$ and $\tr(1)(\Aomega) = 1/3$. 
\end{exa}

Recall, that we still have to prove Proposition~\ref{prop:trace_premeasure}. In order to simplify this proof, we provide a few technical results about the sets $\Sstar$, $\Somega$, $\Sinfty$. For all these results remember again that $\A$ is required to be a \emph{finite} alphabet. This is a crucial point, particularly in the next lemma.

\begin{lem}[Countable Unions]
\label{lem:union_cones}
Let $(S_n)_{n \in \N}$ be a sequence of pairwise disjoint sets in $\Somega$ or in $\Sinfty$ such that their union, $\cup_{n \in \N}S_n$, is itself an element of $\Somega$ or $\Sinfty$. Then $S_n = \emptyset$ for all but finitely many $n$.
\end{lem}
\proof
We have several cases to consider.\\
\emph{Case 1:} If $\cup_{n \in \N}S_n = \emptyset \in \S_\diamond$ for $\diamond \in \{\omega, \infty\}$, we have $S_n = \emptyset$ for all $n \in \N$.\\
\emph{Case 2:} If $\cup_{n \in \N} = \{u\} \in \Sinfty$ with suitable $u \in \Astar$ we get $S_n=\emptyset$ for all but one $n \in \N$ since the $S_n$ are disjoint.\\
\emph{Case 3:} Let $\cup_{n \in \N}S_n =~ \cone{\diamond}{u}$ with a suitable $u \in \Astar$ for $\diamond \in \{\omega, \infty\}$. Suppose there are infinitely many $n \in \N$ such that $S_n \not = \emptyset$. Without loss of generality we can assume $S_n\not=\emptyset$ for all $n \in \N$ and thus there is an infinite set $U :=\set{u_n\mid n \in \N}$ of words such that for each $n \in \N$ we either have $S_n = \{u_n\}$ (only for $\diamond=\infty$) or $S_n =~\cone{\diamond}{u_n}$ (for $\diamond \in \{\omega, \infty\}$). Necessarily we have $u \sqsubseteq u_n$ for all $n \in \N$. We will now revive our tree metaphor from Section \ref{sec:cones}: The prefix-closure  $\mathrm{pref}(U) = \set{v \in \Astar \mid \exists n \in \N: v \sqsubseteq u_n}$ of $U$ is the set of vertices contained in the paths from the root $\epsilon$ (via $u$) to $u_n$. We consider the subtree $\mathcal{T'} = (\mathrm{pref}(U), E', \epsilon, l|_{E'})$ with $E' = \set{\set{u,ua} \mid a \in \A, u, ua \in \mathrm{pref}(U)}$.
Since the set $U$ and hence also $\mathrm{pref}(U)$ is infinite, we have thus
constructed an infinite, connected graph where every vertex has finite degree (because $\A$ is finite). By König's Lemma \cite[Satz 3]{koenigd} there is an infinite path starting at the root $\epsilon$. Let $v  \in \Aomega$ be the unique, infinite word associated to that path (which we get by concatenating all the labels along this path). Since $u \sqsubset v$ we must have $v \in ~\cone{\diamond}{u}$. Moreover, we know that $~\cone{\diamond}{u} = \cup_{n \in \N} S_n$ and due to the fact that the $S_n$ are pairwise disjoint we must find a unique $m \in \N$ with $v \in S_m$. This necessarily requires $S_m$ to be a cone of the form $S_m = ~\cone{\diamond}{u_m}$ with $u_m \in U$ and $u_m \sqsubset v$. Again due to the fact that the $S_n$ are disjoint we know that there cannot be a $u' \in U$ with $u_m \sqsubset u'$ and hence there also cannot be a $u' \in \mathrm{pref}(U)$ with $u_m \sqsubset u'$. Thus the vertex $u_m$ is a leaf of the tree $\mathcal{T}'$ and therefore the finite path from $\epsilon$ to $u_m$ is the only path from $\epsilon$ that contains $u_m$. This contradicts the existence of $v$ because this path is infinite and contains $u_m$. Hence our assumption must have been wrong and there cannot be infinitely many $n \in \N$ with $S_n \not = \emptyset$. 
\qed

\begin{lem}
\label{lem:premeas_singletons}
Any map $\mu \colon \Sstar \to \ExtR_+$ where $\mu(\emptyset) = 0$ is $\sigma$-additive and thus a pre-measure. 
\end{lem}
\proof
Let $\left(S_n\right)_{n \in \N}$ be a family of disjoint sets from $\Sstar$ with $\left(\cup_{n \in \N}S_n\right) \in \Sstar$, then we have $S_n = \emptyset$ for all but at most one $n \in \N$. 
\qed

\begin{lem}
\label{lem:premeas_omega_cone}
A map $\mu\colon \S_\omega \to \ExtR_+$ where $\mu(\emptyset) = 0$ is $\sigma$-additive and thus a pre-measure if and only if the following equation holds for all $u \in \Astar$.
\begin{align}
	\mu\left(\cone{\omega}{u}\right) = \sum_{a \in \A}\mu\left(\cone{\omega}{ua}\right)\label{eq:premeas_omega_cone}
\end{align} 
\end{lem}

\noindent We omit the proof of this lemma as it is very similar to the proof of the following lemma.

\begin{lem}
\label{lem:premeas_infty_cone}
A map $\mu\colon \Sinfty \to \ExtR_+$ where $\mu(\emptyset) = 0$ is $\sigma$-additive and thus a pre-measure if and only if the following equation holds for all $u \in \Astar$. 
\begin{align}
	\mu\left(\cone{\infty}{u}\right) = \mu\left(\set{u}\right) + \sum_{a \in \A}\mu\left(\cone{\infty}{ua}\right)\label{eq:premeas_infty_cone}
\end{align}
\end{lem}
\proof
Obviously $\sigma$-additivity of $\mu$ implies equality \eqref{eq:premeas_infty_cone}. Let now $\left(S_n\right)_{n \in \N}$ be a family of disjoint sets from $\Sinfty$ with $\left(\cup_{n \in \N}S_n\right) \in \Sinfty$. Using Lemma~\ref{lem:union_cones} we know that (after resorting) we can assume that there is an $N \in \N$ such that $S_n \not= \emptyset$ for $1 \leq n \leq N$ and $S_n = \emptyset$ for $n > N$. For non-trivial cases (trivial means $S_n = \emptyset$ for all but one set) there must be a word $u \in \Astar$ such that $\cone{\infty}{u} = \left(\cup_{n = 1}^NS_n\right)$. Because $u$ is an element of $\cone{\infty}{u}$ there must be a natural number $m$ with $u \in S_m$ which is unique because the family is disjoint. Without loss of generality we assume that $u \in S_1$. By construction of $\Sinfty$ and the fact that $\cup_{n=1}^NS_n =~\cone{\infty}{u}$ there are two cases to consider: either $S_1 = \set{u}$ or $S_1=~\cone{\infty}{u}$. The latter cannot be true since this would imply $S_n = \emptyset$ for $n\geq 2$ which we explicitly excluded. Thus we have $S_1 = \set{u}$. We remark that 
\begin{align*}
	\bigcup_{a \in \A} \cone{\infty}{ua} =~\cone{\infty}{u}\setminus \set{u} = \left(\bigcup_{n=2}^NS_n\right).
\end{align*}
Again by construction of $\Sinfty$ we must be able to select sets $S_k^a \in \set{S_n \mid 2 \leq n \leq N}$ 
for all $a \in \A$ and all $k$ where $1 \leq k \leq K_a < N$ for a constant $K_a$ such that $\cup_{k =1}^{K_a} S_k^a =~\cone{\infty}{ua}$. This selection is unique in the following manner: For $a,b\in\A$ where $a \not = b$ and $1\leq k \leq K_a$, $1\leq l \leq K_b$ we have $S_k^a \not= S_l^b$. Additionally it is complete in the sense that $\set{S_k^a\mid a \in \A, 1 \leq k \leq K_a} = \set{S_n\mid 2 \leq n \leq N}$. 
We apply our equation \eqref{eq:premeas_infty_cone} to get
\begin{align*}
	\mu\left(\bigcup_{n=1}^NS_n\right)= \mu\left(\cone{\infty}{u}\right) = \mu\left(S_1\right) + \sum_{a \in \A}\mu\left(\bigcup_{k =1}^{K_a} S_k^a\right)
\end{align*}
and note that we can repeat the procedure for each of the disjoint unions $\cup_{k=1}^{K_a} S_k^a$. Since $K_a < N$ for all $a$ this procedure stops after finitely many steps yielding $\sigma$-additivity of $\mu$. 
\qed

Using these results, we can now finally prove Proposition~\ref{prop:trace_premeasure}.

\proof[Proof of Proposition~\ref{prop:trace_premeasure}]
We will look at the different types of PTS separately. For $\diamond = 0$ nothing has to be shown because $\sigalg[\emptyset]{\set{\emptyset}} = \set{\emptyset}$ and $\tr(x)\colon \set{\emptyset} \to [0,1]$ is already uniquely defined by $\tr(x)(\emptyset) = 0$. For $\diamond = *$ Lemma~\ref{lem:premeas_singletons} yields immediately that the equations define a pre-measure. For $\diamond = \infty$ we have to check validity of equation \eqref{eq:premeas_infty_cone} of Lemma~\ref{lem:premeas_infty_cone}. We will do so using induction on the length of the word $u \in \Astar$ in that equation. We have
\begin{align*}
	&\tr(x)(\cone{\infty}{\epsilon}) = 1 = \alpha(x)(\A \times X + \one) = \alpha(x)(\one) + \sum_{a \in \A}\P{a}{x}{X} \\
		&= \tr(x)(\set{\epsilon}) + \sum_{a \in \A}\Int[x'\in X]{1}[\P{a}{x}{x'}]\\
		&= \tr(x)(\set{\epsilon}) + \sum_{a \in \A}\Int[x'\in X]{\tr(x')(\cone{\infty}{\epsilon})}[\P{a}{x}{x'}]\\
		&=\tr(x)(\set{\epsilon}) + \sum_{a \in \A}\tr(x)(\cone{\infty}{a\epsilon}) = \tr(x)(\set{\epsilon}) + \sum_{a \in \A}\tr(x)(\cone{\infty}{\epsilon a})
\end{align*}
for all $x \in X$. Now let us assume that for all $x \in X$ and all words $u \in \A^{\leq{n}}$ of length less or equal to a fixed $n \in \N$ the induction hypothesis
\begin{align*}
	\tr(x)(\cone{\infty}{u}) = \tr(x)(\set{u}) + \sum_{b \in \A} \tr(x)(\cone{\infty}{ub})
\end{align*}
is fulfilled. Then for all $x \in X$, all $a \in \A$ and all $u \in \A^{\leq n}$ we calculate
\begin{align*}
	&\tr(x)(\cone{\infty}{au}) = \Int[{x' \in X}]{\tr(x')(\cone{\infty}{u})}[\P{a}{x}{x'}] \\
		&\quad = \Int[{x' \in X}]{\left(\tr(x')(\set{u}) + \sum_{b \in \A} \tr(x')(\cone{\infty}{ub})\right)}[\P{a}{x}{x'}]\\
		&\quad = \Int[{x' \in X}]{\tr(x')(\set{u})}[\P{a}{x}{x'}] + \sum_{b \in \A} \Int[x'\in X]{\tr(x')(\cone{\infty}{ub})}[\P{a}{x}{x'}]\\
		&\quad = \tr(x)(\set{au}) + \sum_{b \in \A}\tr(x)(\cone{\infty}{aub})
\end{align*}
and hence also for $au \in \A^{\leq {n+1}}$ equation \eqref{eq:premeas_infty_cone} is fulfilled and by induction we conclude that it is valid for all $u \in \Astar$. The only difficult case is $\diamond = \omega$ where we will, of course, apply Lemma~\ref{lem:premeas_omega_cone}. Let $u = u_1\dots u_m$ with $u_k \in \A$ for every $k \in \N$ with $k \leq m$, then multiple application of the defining equation \eqref{eq:trace_main_equation} yields
\begin{align*}
	\tr(x)\big(\cone{\omega}{u}\big) &= \int\limits_{x_1 \in  X}\hdots\int\limits_{x_m \in X}\!1\,\mathrm{d}\P{u_m}{x_{m-1}}{x_m}\hdots\mathrm{d}\P{u_1}{x}{x_1}
\end{align*}
and for arbitrary $a \in \mathcal{A}$ we obtain analogously:
\begin{align*}
	\tr(x)\big(\cone{\omega}{ua}\big) &= \int\limits_{x_1 \in X}\hdots\int\limits_{x_m \in X}\!\P{a}{x_m}{X}\,\mathrm{d}\P{u_m}{x_{m-1}}{x_m}\hdots\mathrm{d}\P{u_1}{x}{x_1}\,.
\end{align*}
All integrals exist and are bounded above by $1$ so we can use the linearity and monotonicity of the integral to exchange the finite sum and the integrals. Using the fact that
\begin{align*}
	\sum_{a \in \mathcal{A}}\P{a}{x_m}{X} = \sum_{a \in \mathcal{A}}\alpha(x_m)(\set{a} \times X) = \alpha(x_m)(\A \times X) = 1
\end{align*}
we obtain that indeed the necessary and sufficient equality
\begin{align*}
	 \tr(x)\big(\cone{\omega}{u}\big) = \sum_{a \in \mathcal{A}}\tr(x)\big(\cone{\omega}{ua}\big)
\end{align*} is valid for all $u \in \mathcal{A}^*$ and thus Lemma~\ref{lem:premeas_omega_cone} yields that also $\tr(x)\colon \Somega \to \ExtR_+$ is $\sigma$-additive and thus a pre-measure.

Now let us check that the pre-measures for $\diamond \in \set{*, \omega, \infty}$ are $\sigma$-finite and that their unique extensions must be (sub-)probability measures. For $\diamond \in \set{\omega, \infty}$ this is obvious and in these cases the unique extension must be a probability measure because by definition we have $\tr(x)(\Aomega) = 1$ and $\tr(x)(\Ainfty) = 1$ respectively. For the remaining case ($\diamond = *$) we will use induction. We have $\tr(x)(\{\epsilon\}) = \alpha(x)(\one) \leq 1$ for every $x \in X$. Let us now assume that for a fixed but arbitrary $n \in \N$ the inequality $\tr(x)(\{u\}) \leq 1$ is valid for all $x \in X$ and all words $u \in \mathcal{A}^{\leq n}$ with length less or equal to $n$. Then for any word $u' \in \mathcal{A}^{n+1}$ of length $n+1$ we have $u' = au$ with $a \in \mathcal{A}$ and $u \in \mathcal{A}^n$. We observe that
\begin{align*}
	\tr(x)(\{au\}) = \Int[{x' \in  X}]{\underbrace{\tr(x')(\{u\})}_{\leq  1}}[\P{a}{x}{x'}] \leq \Int{1}[\P{a}{x}{x'}]=\P{a}{x}{X} \leq 1
\end{align*}
and conclude by induction that $\tr(x)(\{u\}) \leq 1$ is valid for all $u \in \mathcal{A}^*$ and all $x \in X$. Due to the fact that $\mathcal{A}^* = \cup_{u \in \mathcal{A}^*} \{u\}$ this yields that $\tr$ is $\sigma$-finite.

Again by induction we will show that $\tr$ is bounded above by $1$ and thus a sub-probability measure. We have $\tr(x)\left(\mathcal{A}^{\leq 0}\right) = \tr(x)(\{\epsilon\})\leq 1$ for all $x \in X$. Suppose that for a fixed but arbitrary  $n \in \mathbb{N}$ the inequality $\tr(x)\left(\mathcal{A}^{\leq n-1}\right) \leq 1$ holds for all $x \in X$. We conclude with the following calculation
\begin{align*}
	\tr(x)\left(\mathcal{A}^{\leq n}\right) &= \tr(x)\left( \cup_{u \in \mathcal{A}^{\leq n}} \{u\}\right) = \sum\limits_{u \in \mathcal{A}^{\leq n}} \tr(x)\left( \{u\}\right)\\
	&= \tr(x)(\{\epsilon\}) + \sum_{a \in \mathcal{A}} \sum_{u  \in \mathcal{A}^{\leq n-1}} \tr(x)\left(\{au\}\right) \\
	&= \alpha(x)(\one) + \sum\limits_{a \in \mathcal{A}} \sum\limits_{u  \in \mathcal{A}^{\leq n-1}} \Int{\tr(x')\left(\set{u}\right)}[\P{a}{x}{x'}] \\
	&= \alpha(x)(\one) + \sum\limits_{a \in \mathcal{A}} \Int{\sum_{u  \in \mathcal{A}^{\leq n-1}}\!\tr(x')(\{u\})}[\P{a}{x}{x'}]\\
	&= \alpha(x)(\one) + \sum\limits_{a \in \mathcal{A}} \Int{\underbrace{\left(\tr(x')\left(\mathcal{A}^{\leq n-1}\right)\right)}_{\leq 1}}[\P{a}{x}{x'}]\\
	&\leq \alpha(x)(\one) + \sum\limits_{a \in \mathcal{A}}\Int{1}[\P{a}{x}{x'}]=\alpha(x)(\one) + \sum\limits_{a \in \mathcal{A}} \P{a}{x}{X}\\
	&= \alpha(x)(\one) + \sum\limits_{a \in \mathcal{A}} \alpha(x)(\{a\} \times X)= \alpha(x)(\mathcal{A} \times X + \one) \leq 1
\end{align*}
using the linearity and monotonicity of the integral which can be applied here since $\mathcal{A}$ is finite which in turn implies that $\mathcal{A}^{\leq n-1}$ is finite and all the integrals $\Int{\tr(x')\left(\set{u}\right)}[\P{a}{x}{x'}]$ exist because $\tr(x')\left(\set{u}\right)$ is bounded above by $1$. By induction we can thus conclude that
\begin{align*}
	\forall x \in X\ \forall n \in \mathbb{N}_0: \tr(x)\left(\mathcal{A}^{\leq n} \right) \leq 1
\end{align*}
which is equivalent to
\begin{align*}
	\forall x \in X\ \sup_{n \in \mathbb{N}_0}\left(\tr(x)\left(\mathcal{A}^{\leq n} \right)\right) \leq 1\,.
\end{align*}
Since $\tr(x)$ is a measure (and thus non-negative and $\sigma$-additive), the sequence given by $\left(\tr(x)\left(\mathcal{A}^{\leq n}\right)\right)_{n \in  \mathbb{N}_0}$ is a monotonically increasing sequence of real numbers bounded above by $1$. Furthermore, $\tr(x)$ is continuous from below as a measure and we have $\mathcal{A}^{\leq n} \subseteq \mathcal{A}^{\leq n+1}$ for all $n \in \N_0$ and thus we obtain
\begin{align*}
	\tr(x)\left(\mathcal{A}^*\right) = \tr(x)\left( \bigcup\limits_{n =1}^\infty \mathcal{A}^{\leq n}\right)  = \lim_{n \to \infty} \tr(x)\left( \mathcal{A}^{\leq n}\right) = \sup_{n \in\N_0}\tr(x)\left(\mathcal{A}^{\leq n} \right) \leq 1\,.
\end{align*}
\qed

\subsection{The Trace Function is a Kleisli Arrow}
Now that we know that our definition of a trace measure is mathematically sound, we remember that we wanted to show that it is ``natural'', meaning that it arises from the final coalgebra in the Kleisli category of the (sub-)probability monad. We start by showing that the function $\tr\colon X \to T\A^\diamond$ we get from Definition \ref{def:trace_premeasure} is a Kleisli arrow by proving that it is a Markov kernel. Since  $\tr(x)$ is a sub-probability measure for each $x \in X$ by Proposition \ref{prop:trace_premeasure} we just have to show that for each $S \in \sigalg[\A^\diamond]{\S_\diamond}$ the function $x \mapsto \tr(x)(S)$ is Borel-measurable. This is easy for elements $S$ of the previously defined semirings:

\begin{lem}
	\label{lem:measurabilityGenerators}
	For every $S \in \S_\diamond$ the function $x \mapsto \tr(x)(S)$ is Borel-measurable.
\end{lem}
\proof
For $\diamond=0$ nothing has to be shown. For the other cases we will use induction on the length of a word $u$. For $\diamond \in \set{*,\infty}$ measurability of $x \mapsto \tr(x)(\set{\epsilon})$ follows from measurability of $x \mapsto \alpha(x)(\one)$ and for $\diamond \in \set{\omega,\infty}$ the function $x \mapsto \tr\left(x)(\cone{\diamond}{\epsilon}\right)$	is the constant function with value $1$ and thus is measurable. Suppose now that for an $n \in \N$ we have established that for all $u \in \A^n$ the functions $x \mapsto \tr(x)(\set{u})$ and $x \mapsto \tr(x)(\cone{\diamond}{u})$ (whenever they are meaningful) are measurable. Then for all $a \in \A$ and all $u \in \A^n$ we have $\tr(x)(\set{au}) = \Int[x' \in X]{\tr(x')(\set{u})}[{\mathbf{P}_{a}(x,x')}]$ and also $\tr(x)(\cone{\diamond}{au}) = \Int[x' \in X]{\tr(x')(\cone{\diamond}{u})}[{\mathbf{P}_{a}(x,x')}]$ and by applying Lemma \ref{lem:measMarkovKernel} we get the desired measurability.
\qed

Without any more complicated tools we get the complete result for any $*$-PTS:

\begin{prop}
	\label{prop:MeasurabilityFiniteTrace}
	 For every $S \in \powerset{\Astar}$ the function $x \mapsto\tr(x)(S)$ is Borel-measurable.
\end{prop}
\proof
We know from Lemma \ref{lem:measurabilityGenerators} that $x\mapsto\tr(x)(S)$ is measurable for every $S \in \Sstar$. Recall that $\sigalg[\Astar]{\Sstar} = \powerset{\Astar}$ and every $S \in \powerset{\Astar}$ is at most countably\footnote{For finite $S$ the proof works analogously but simpler!} infinite, say $S := \set{u_1, u_2,\hdots}$ and we have the trivial, disjoint decomposition $S = \cup_{n=1}^\infty\set{u_n}$. If we define $T_N := \cup_{n=1}^N \set{u_n}$ we get an increasing sequence of sets converging to $S$. Hence by continuity of the sub-probability measures $S' \mapsto \tr(x)(S')$ we get $\tr(x)(S) = \lim_{N\to \infty}\tr(x)(T_N) = \lim_{N\to \infty}\sum_{n=1}^N \tr(x, \set{u_n})$. Thus $x \mapsto\tr(x)(S)$ is the pointwise limit of a finite sum of measurable functions and therefore measurable.
\qed

From here until the rest of this subsection we restrict $\diamond$ to be either $\omega$ or $\infty$ if not indicated otherwise. As before, we will rely on transfinite induction for our proof.

\begin{lem}
	For every $S \in \Rdiamond(0)$ the function $x \mapsto\tr(x)(S)$ is measurable.
\end{lem}
\proof
	We know from Lemma \ref{lem:measurabilityGenerators} that $x\mapsto\tr(x)(S)$ is measurable for every $S \in \S_\diamond$. Let $S \in \Rdiamond(0)$ then $S = \cup_{n=1}^N S_n$ with $S_n \in \S_\diamond$ disjoint for $1 \leq n \leq N \in \N$. We have $\tr(x)(S) = \sum_{n=1}^N \tr(x,S_n)$ which is measurable as a finite sum of measurable functions.
\qed

\begin{lem}
	Let $\alpha$ be an ordinal s.t. the function $x \mapsto \tr(x)(S)$ is measurable for each $S \in \Rdiamond(\alpha)$. Then $x \mapsto \tr(x)(S)$ is measurable for each $S \in \Intersections{\Rdiamond(\alpha)}$.
\end{lem}
\proof
	Let $S \in \Intersections{\Rdiamond(\alpha)}$ then $S = \cap_{n=1}^\infty S_n$ with $S_n \in \Rdiamond(\alpha)$. We define $T_N := \cap_{n=1}^N S_n$ for all $N \in \N$, then $T_N \in \Rdiamond(\alpha)$ by Lemma \ref{lem:finite_union_intersection}. We have $T_N \supseteq T_{N+1}$ for all $N \in \N$ \linebreak and $S = \cap_{N=1}^\infty T_N$. Continuity of $S' \mapsto \tr(x)(S')$ for every $x \in X$ yields $\tr(x)(S) = \lim_{N \to \infty} \tr(x)\left(T_N\right)$. Hence $x \mapsto \tr(x)(S)$ is measurable as pointwise limit of measurable functions.
\qed

\begin{lem}
	Let $\alpha$ be an ordinal s.t. the function $x \mapsto \tr(x)(S)$ is measurable for each $S \in \Intersections{\Rdiamond(\alpha)}$. Then $x \mapsto \tr(x)(S)$ is measurable for each $S \in \Rdiamond(\alpha+1)$.
\end{lem}
\proof
Let $S \in \Rdiamond(\alpha+1)$ then $S = \cup_{n=1}^\infty S_n$ with $S_n \in \Intersections{\Rdiamond(\alpha)}$. We define $T_N:= \cup_{n=1}^N S_n$ for all $N \in \N$. Then we know that $T_N \in \Intersections{\Rdiamond(\alpha)}$ by Lemma \ref{lem:IntersectionIsClosedUnderFiniteUnion}. We have \linebreak $T_N \subseteq T_{N+1}$ for all $N \in \N$ and $S = \cup_{N=1}^\infty T_N$. Continuity of the sub-probability measures $S' \mapsto \tr(x)(S')$ yields for every $x \in X$ that $\tr(x)(S) = \lim_{N \to \infty} \tr\left(x)(T_N\right)$. Hence the function $x \mapsto \tr(x)(S)$ is measurable as pointwise limit of measurable functions.
\qed

\begin{lem}
	Let $\gamma$ be a limit ordinal s.t. for all ordinals $\alpha < \gamma$ the function $x \mapsto \tr(x)(S)$ is measurable for each $S \in \Rdiamond(\alpha)$. Then $x \mapsto \tr(x)(S)$ is measurable for each $S \in \Rdiamond(\gamma)$.
\end{lem}
\proof
Let $S \in \Rdiamond(\gamma)$, then there is an $\alpha < \gamma$ such that $S \in \Rdiamond(\alpha)$ and hence $x \mapsto\tr(x)(S)$ is measurable for this $S$.
\qed
By using the characterization $\sigalg[\A^\diamond]{\S_\diamond} = \Rdiamond(\omega_1)$ of Proposition \ref{prop:TransFiniteSigAlg} and combining the four preceding lemmas we get the desired result:
\begin{prop}
	\label{prop:TraceIsMeasurable}
	For every $S \in \sigalg[\A^\diamond]{\S_\diamond}$ the function $x \mapsto \tr(x)(S)$ is measurable. \qed
\end{prop}

Finally, combining this result with Proposition \ref{prop:trace_premeasure} and the fact that Markov kernels are in one-to-one correspondence with Kleisli arrows \cite[Proposition 2.7]{Dob07b} yields:

\begin{prop}
	\label{prop:TraceIsKleisli}
	Let $\diamond \in \set{0, *, \omega, \infty}$ and $(T, \eta, \mu)$ be the (sub-)probability monad. Then the function $\tr\colon X \to T\A^\diamond$ given by Definition \ref{def:trace_premeasure} is a Kleisli arrow. \qed
\end{prop}

\subsection{The Trace Measure and Final Coalgebra}
Before stating the next proposition which presents a close connection between the unique existence of the map into the final coalgebra and the unique extension of a family of $\sigma$-finite pre-measures, we first give some intuition: in order to show that a coalgebra is final it is enough to show that every other coalgebra admits a unique homomorphism into it. Commutativity of the square underlying the homomorphism and uniqueness have to be shown for every element of a $\sigma$-algebra and one of our main contributions is to reduce the proof obligations to a smaller set of generators, which form a covering semiring. This proposition will later be applied to our four types of transition systems by using the semirings defined earlier and showing that there can be only one way to assign probabilities to their elements.\newpage

\begin{prop}
\label{thm_finalcoalg}
Let $(T, \eta, \mu)$ be either the sub-probability monad $(\mathbb{S}, \eta, \mu)$ or the probability monad $(\mathbb{P}, \eta, \mu)$, $F$ be an endofunctor on $\Meas$ with a distributive law $\lambda \colon FT \Rightarrow TF$ and $(\Omega, \kappa)$ be an $\overline{F}$-coalgebra where $\Sigma_{F\Omega} = \sigalg[F\Omega]{\S_{F\Omega}}$ for a covering semiring $\S_{F\Omega}$. Then the following statements are equivalent:
\begin{enumerate}
	\item $(\Omega, \kappa)$ is a final $\overline{F}$-coalgebra in $\Kl(T)$.
	\item For every $\overline{F}$-coalgebra $(X, \alpha)$ in $\Kl(T)$ there is a unique Kleisli arrow $\tr\colon X \to T\Omega$ satisfying the following condition:
\begin{align}
	\forall x \in X, \forall S \in \S_{F\Omega}: \quad \Int[\Omega]{p_S \circ \kappa}[\tr(x)] = \Int[{FX}]{p_S \circ \lambda_\Omega \circ F(\tr)}[\alpha(x)]\,. \label{eq:giry_final_coalgebra}
\end{align}
\end{enumerate}
\end{prop}
\proof
We consider the final coalgebra diagram in $\Kl(T)$.
\begin{align*}\begin{xy}\xymatrix{
 	X  \ar[rr]^{\alpha} \ar[d]_{\tr} &&  \overline{F}X  \ar[d]^{\overline{F}(\tr) = \lambda_\Omega \circ F(\tr)}\\
	\Omega  \ar[rr]^{\kappa} &&  \overline{F}\Omega
}\end{xy}\end{align*}
By definition $(\Omega, \kappa)$ is final iff for every $\overline{F}$-coalgebra $(X , \alpha)$ there is a unique Kleisli arrow $\tr \colon X  \to T\Omega$ making the diagram commute. We define
 \begin{align*}
 	g := \mu_{F\Omega} \circ T(\kappa)\circ \tr\ \mbox{(down, right)} \quad \text{and}\quad h:= \mu_{F\Omega} \circ T\left(\overline{F}(\tr)\right) \circ \alpha\ \mbox{(right, down)} 
\end{align*}
and note that commutativity of the final coalgebra diagram is equivalent to
\begin{align}
	\forall x \in X,\forall S \in \S_{F\Omega}: \quad g(x)(S) &= h(x)(S) \label{eq:g_equals_h}
\end{align}
because $\S_{F\Omega}$ is a covering semiring and for all $x \in X$ both $g(x)$ and $h(x)$ are sub-probability measures and thus finite measures which allows us to apply Corollary \ref{cor:equality_of_measures}. We calculate
\begin{align*}
	g(x) (S) 	&= (\mu_{F\Omega} \circ T(\kappa) \circ \tr) (x) (S)  =\mu_{F\Omega}\left(T(\kappa)(\tr(x))\right)(S)\\
			& =\mu_{F\Omega} \left(\tr(x) \circ \kappa^{-1}\right) (S) = \Int{p_S}[{\left(\tr(x) \circ \kappa^{-1}\right)}] = \Int{p_S\circ\kappa} [\tr(x)]
\end{align*}
and if we define $\rho := \overline{F}(\tr) = \lambda_\Omega \circ F(\tr) \colon FX \to TF\Omega$ we obtain
\begin{align*}
	h(x)(S) 	&= (\mu_{F\Omega} \circ T(\rho) \circ \alpha) (x) (S) = \mu_{F\Omega} \left(T(\rho) (\alpha (x))\right) (S) =\mu_{F\Omega} \left(\alpha(x)\circ \rho^{-1}\right) (S) \\
			&= \int\! p_S\, \mathrm{d}\left(\alpha(x)\circ \rho^{-1}\right) =  \int\! p_S \circ \rho\, \mathrm{d}\alpha(x) = \int\! p_S \circ \lambda_\Omega \circ F(\tr)\, \mathrm{d}\alpha(x) 
\end{align*}
and thus \eqref{eq:g_equals_h} is equivalent to \eqref{eq:giry_final_coalgebra}.
\qed

We immediately obtain the following corollary.

\begin{cor}
\label{cor:main_corollary}
Let in Proposition \ref{thm_finalcoalg} $\kappa = \eta_{F\Omega} \circ \phi$, for an isomorphism $\phi \colon \Omega \to F\Omega$ in $\Meas$, and let $\S_\Omega \subseteq \powerset{\Omega}$ be a covering semiring such that $\Sigma_\Omega = \sigalg[\Omega]{\S_\Omega}$. Then equation \eqref{eq:giry_final_coalgebra} is equivalent to
\begin{align}
	\forall x \in X, \forall S \in \S_\Omega: \quad \tr(x)(S) = \Int{p_{\phi(S)} \circ \lambda_\Omega \circ F(\tr)}[\alpha(x)]\,. \label{eq:giry_final_coalgebra_semiring}
\end{align}
\end{cor}
\proof
Since $\phi$ is an isomorphism in $\Meas$ we know from Proposition~\ref{prop:isomorphisms} that $\Sigma_{F\Omega} = \sigalg[F\Omega]{\phi(\S_\Omega)}$. For every $S \in \S_{\Omega}$ and every $u \in \Omega$ we calculate
\[p_{\phi(S)}\circ \kappa (u) = p_{\phi(S)}\ \circ\eta_{F\Omega} \circ \phi (u) = \delta_{\phi(u)}^{F\Omega} (\phi(S))= \chi_{\phi(S)}(\phi(u)) = \chi_{S}({u}) \] 
and hence we have $\int\! p_{\phi(S)}\circ\kappa\, \mathrm{d}\tr(x) = \int\! \chi_S \, \mathrm{d}\tr(x) = \tr(x)(S)$. 
\qed

Since we want to apply this corollary to sets of words, we now define the necessary isomorphism $\phi$ using the characterization given in Proposition~\ref{prop:isomorphisms}. 

\begin{prop}
\label{prop:words_iso}
Let $\mathcal{A}$ be an arbitrary alphabet and let 
\begin{align}
	\phi\colon \Ainfty \to \A \times \Ainfty + \one, \quad \epsilon \mapsto \checkmark, \quad au \mapsto (a,u)\,.
\end{align}
Then $\phi$, $\phi|_{\Astar}\colon \Astar \to \phi(\Astar)$ and $\phi|_\Aomega\colon \Aomega \to \phi(\Aomega)$ are isomorphisms in $\Meas$ because they are bijective functions\footnote{Note that we restrict not only the domain of $\phi$ here but also its codomain.} and we have
\begin{align}
	\sigalg[\A \times \Aomega]{\phi(\Somega)} &= \powerset{\A} \otimes \sigalg[\Aomega]{\Somega}\,, \\
	\sigalg[\A \times \Astar + \one]{\phi(\Sstar)} &= \powerset{\A} \otimes \sigalg[\Astar]{\Sstar} \oplus \powerset{\one}\,, \label{eq:sigalg_eq_Astar}\\
	\sigalg[\A \times \Ainfty + \one]{\phi(\Sinfty)} &= \powerset{\A} \otimes \sigalg[\Ainfty]{\Sinfty} \oplus \powerset{\one}\,. \label{eq:sigalg_eq_Ainfty}
\end{align}
\end{prop}
\proof
Bijectivity is obvious. We will now show validity of \eqref{eq:sigalg_eq_Ainfty}, the other equations can be verified analogously.\footnote{For proving \eqref{eq:sigalg_eq_Astar} we can use Proposition \ref{prop:generator_product} because $\sigalg[\Astar]{\Sstar} = \sigalg[\Astar]{\Sstar \cup \set{\Astar}}$.} Let $\S_\A := \set{\emptyset} \cup \set{\set{a} \mid a \in \A} \cup \set{\A}$, then it is easy to show that we have $\sigalg[\A]{\mathcal{S}_\A} = \powerset{\A}$ and Propositions \ref{prop:generator_product} and \ref{prop:generator_union} yield that 
\begin{align*}
	\powerset{\A} \otimes \sigalg[\Ainfty]{\Sinfty} \oplus \powerset{\one} = \sigalg[\A \times \Ainfty + \one]{\S_\A \ast \Sinfty \oplus \powerset{\one}}\,.
\end{align*}
We calculate $\phi\left(\emptyset\right) = \emptyset$, $\phi\left(\set{\epsilon}\right) = \one$, $\phi\left(\cone{\omega}{\epsilon}\right) = \A \times \Aomega$, $\phi\left(\cone{\infty}{\epsilon}\right) = \A \times \Ainfty + \one$, and for all $a \in \A$ and all $u \in \Astar$ we have $\phi\left(\set{au}\right) = \set{(a,u)}$ and also $\phi\left(\cone{\infty}{au}\right) = \set{a} \times \cone{\infty}{u}$. This yields
\begin{align*}
	\phi(\Sinfty) &= \set{\emptyset, \emptyset + \one, \A \times \Ainfty + \one} \cup \set{\set{a} \times \set{u} + \emptyset, \set{a} \times \cone{\infty}{u}+\emptyset\mid a \in \A, u \in \Astar}
\end{align*}
and furthermore we have
\begin{align*}
	\mathcal{S}_\A \ast \Sinfty \oplus \powerset{\one} = \set{\emptyset, \emptyset + \one} &\cup \set{\set{a} \times \set{u} + \emptyset, \set{a} \times \cone{\infty}{u}+\emptyset\mid a \in \A, u \in \Astar}\\
	&\cup \set{\set{a} \times \set{u} + \one, \set{a} \times \cone{\infty}{u}+\one\mid a \in \A, u \in \Astar}\\
	&\cup \set{\A\, \times \set{u} + \emptyset, \A\, \times \cone{\infty}{u}+\emptyset\mid u \in \Astar}\\
	&\cup \set{\A\, \times \set{u} + \one, \A\, \times \cone{\infty}{u}+\one\mid u \in \Astar}.
\end{align*}
Due to the fact that $\A \times \Ainfty + \one = \A \times \cone{\infty}{\epsilon} + \one$ we have $\phi(\Sinfty) \subseteq \S_\A \ast \Sinfty \oplus \powerset{\one}$ and the monotonicity of the $\sigma$-operator yields 
\begin{align*}
	\sigalg[\A \times \Ainfty + \one]{\phi(\Sinfty)} \subseteq \sigalg[\A \times \Ainfty + \one]{\S_\A \ast \Sinfty \oplus \powerset{\one}}\,.
\end{align*}
For the other inclusion we remark that 
\begin{align*}
	\set{a} \times \set{u} + \one &= (\set{a} \times \set{u} + \emptyset) \cup (\emptyset + \one)\\
	\set{a} \times \cone{\infty}{u} + \one &= (\set{a} \times \cone{\infty}{u} + \emptyset) \cup (\emptyset + \one)
\end{align*}
and together with the countable decomposition $\A = \cup_{a \in A} \set{a}$ it is easy to see that 
\begin{align*}
	\S_\A \ast \Sinfty \oplus \powerset{\one} \subseteq \sigalg[\A \times \Ainfty + \one]{\phi(\Sinfty)}
\end{align*}
and monotonicity and idempotence of the $\sigma$-operator complete the proof.
\qed

We recall that -- in order to get a lifting of an endofunctor on $\Meas$ -- we also need a distributive law for the functors and the monads we are using to define PTS. In order to define such a law we first provide two technical lemmas.
\begin{lem}
\label{lem:semirings}
Let $\A$ be an alphabet and $(X, \Sigma_X)$ be a measurable space. 
\begin{enumerate}
	\item The sets $\powerset{\A} \ast \Sigma_X$ and $\powerset{\A} \ast \Sigma_X \oplus \powerset{\one}$ are covering semirings of sets.\label{itm:semirings:one}
	\item $\powerset{\A} \otimes \Sigma_X = \sigalg[\A \times X]{\powerset{\A} \ast \Sigma_X}$.
	\item $\powerset{\A} \otimes \Sigma_X \oplus \powerset{\one} = \sigalg[\A \times X+\one]{\powerset{\A} \ast \Sigma_X\oplus \one}$.
\end{enumerate}
\end{lem}
\proof 
Showing property \eqref{itm:semirings:one} is straightforward and will thus be omitted. The rest follows by Propositions \ref{prop:generator_product} and \ref{prop:generator_union}.
\qed

\begin{lem}[Product Measures]
\label{lem:productmeasures}
Let $\A$ be an alphabet, $a \in \A$ and $(X, \Sigma_X)$ be a measurable space with a sub-probability measure $P \colon \Sigma_X \to [0,1]$. Then the following holds:
\begin{enumerate}
	\item The \emph{product measure} $\delta_a^\A \otimes  P\colon \powerset{\A} \otimes \Sigma_X \to \R_+$ of $\delta_a^\A$ and $P$ which is the unique extension of the pre-measure satisfying
	\begin{align}
		 (\delta_a^\A \otimes P)(S_\A \times S_X) := \delta_a^\A(S_\A) \cdot P(S_X)\label{eq:product_measure}
	\end{align}
	for all $S_\A \times S_X \in \powerset{\A} \ast \Sigma_X$ is a sub-probability measure on $\A \times X$. If $P$ is a probability measure on $X$, then also $\delta_a^\A \otimes  P$ is a probability measure on $\A \times X$.
	\item The measure $\delta_a^\A \odot  P\colon \powerset{\A} \otimes \Sigma_X \oplus \powerset{\one} \to \R_+$ which is defined via the equation
	\begin{align}
		\quad (\delta_a^\A \odot  P)(S) := (\delta_a^\A \otimes P)  \left(S \cap (\A \times X)\right)\label{eq:product_coproduct_measure}
	\end{align} 
	for all $S \in \powerset{\A} \otimes \Sigma_X \oplus \powerset{\one}$ is a sub-probability measure on $\A \times X + \one$. If $P$ is a probability measure on $X$, then also $\delta_a^\A \odot  P$ is a probability measure on $\A \times X+\one$.
\end{enumerate} 
\end{lem}
\proof
Before proving the statement, we check that the two equations yield unique measures.
\begin{enumerate}
	\item Existence and uniqueness of the product measure is a well known fact from measure theory and follows immediately by Proposition~\ref{prop:extension} because equation \eqref{eq:product_measure} defines a $\sigma$-finite pre-measure on $\powerset{A} \ast \Sigma_X$ which by Lemma~\ref{lem:semirings} is a covering semiring of sets and a generator for the product-$\sigma$-algebra. 
	\item We obviously have non-negativity and $(\delta_a^\A \odot P)(\emptyset)=0$. Let $(S_n)_{n \in \N}$ be a family of pairwise disjoint sets in $\powerset{A} \otimes \Sigma_X \oplus \powerset{\one}$. Then the following holds
		\begin{align*}
			&(\delta_a^\A \odot  P)\left(\bigcup_{n \in \N}S_n\right) = (\delta_a^\A \otimes  P)\left(\bigcup_{n \in \N}(S_n\cap (\A \times X))\right)\\
			&\quad =\sum_{n \in N}(\delta_a^\A \otimes  P)(S_n\cap (\A \times X)) =\sum_{n \in N}(\delta_a^\A \odot  P)\left(S_n\right)
	\end{align*}
	and hence $\delta_a^\A \odot P$ as defined by equation \eqref{eq:product_coproduct_measure} is $\sigma$-additive and thus a measure.
\end{enumerate}
For the proof of the Lemma we observe that
\begin{align*}
	(\delta_a^\A \odot  P)(\A \times X+\one) = (\delta_a^\A \otimes P)(\A \times X) =  \delta_a^\A(\A) \cdot P(X) = P(X)
\end{align*}
which immediately yields that both measures are sub-probability measures and if $P$ is a probability measure they are probability measures.
\qed

With the help of the preceding lemmas, we can now state and prove the distributive laws for the endofunctors $\A \times \Id_\Meas$, $\A \times \Id_\Meas + \one$ on $\Meas$ and the sub-probability monad and the probability monad.

\begin{prop}[Distributive Laws]
\label{prop:distributive_law_giry}
Let $(T, \eta, \mu)$ be either the sub-probability monad $(\mathbb{S}, \eta, \mu)$ or the probability monad $(\mathbb{P}, \eta, \mu)$ and $\A$ be an alphabet with $\sigma$-algebra $\powerset{\A}$.
\begin{enumerate}
\item Let $F = \mathcal{A} \times \Id_\Meas$. For every measurable space $(X, \Sigma_X)$ we define
\begin{align}
	\lambda_X&\colon \A \times TX  \to T(\A \times X),~ (a,P) \mapsto \delta_a^\A \otimes  P\,.\label{eq:distributivelaw_noterm}
\end{align}
Then $\lambda\colon FT \Rightarrow TF$ is a distributive law.
\item Let $F = \mathcal{A} \times \Id_\Meas + \one$. For every measurable space $(X, \Sigma_X)$ we define
\begin{align}
	&\lambda_X\colon \A \times TX + \one \to T(\A \times X + \one)\nonumber\\
	&(a,P) \mapsto \delta_a^\A \odot  P \label{eq:distributivelaw_term}, \quad \checkmark \mapsto \delta_\checkmark^{\A \times X + \one}\, .
\end{align}
Then $\lambda\colon FT \Rightarrow TF$ is a distributive law.
\end{enumerate}
\end{prop}
\proof
In order to show that the given maps are distributive laws we have to check commutativity of the following three diagrams
\[\xymatrix{
FTY \ar[r]^{\lambda_Y} \ar[d]_{FTf} & TFY \ar[d]^{TFf} & FX \ar[r]^{F\eta_X} \ar[dr]_{\eta_{FX}} & FTX \ar[d]^{\lambda_X} & FT^2X \ar[r]^{\lambda_{TX}} \ar[d]_{F\mu_X} & TFTX \ar[r]^{T\lambda_X} & T^2FX \ar[d]^{\mu_{FX}}\\
FTX \ar[r]^{\lambda_X} & TFX & & TFX & FTX\ar[rr]_{\lambda_X} & & TFX
}\]
for all measurable spaces $(X, \Sigma_X)$, $(Y, \Sigma_Y)$ and all measurable functions $f \colon Y \to X$. By Lemma~\ref{lem:semirings} we know that $\powerset{\A} \ast \Sigma_X$ and $\powerset{\A} \ast \Sigma_X \oplus \powerset{\one}$ are covering semirings of sets and that they are generators for the $\sigma$-algebras $\powerset{\A} \otimes \Sigma_X$ and $\powerset{\A} \otimes \Sigma_X\oplus \powerset{\one}$. Moreover, we know from Lemma~\ref{lem:productmeasures} that the measures assigned in equations~\eqref{eq:distributivelaw_noterm} and \eqref{eq:distributivelaw_term} are sub-probability measures and thus finite. We can therefore use Corollary \ref{cor:equality_of_measures} to check the equality of the various (sub-)probability measures.  
We will provide the proofs for the second distributive law only, the proofs for the first law are simpler and can in fact be derived from the given proofs. Let $S := S_\mathcal{A} \times S_X + S_\one \in \powerset{\A} \ast \Sigma_X \oplus \powerset{\one}$. 
\begin{enumerate}
\item Let $f \colon Y \to X$ be a measurable function. For $(a,P) \in \mathcal{A} \times TY$ we calculate
\begin{align*}
	(TFf \circ \lambda_Y)(a,P)(S) &= (\delta_a^{\mathcal{A}}\odot P)\left((Ff)^{-1}(S)\right) =(\delta_a^{\mathcal{A}}\odot P) (S_{\mathcal{A}} \times f^{-1}(S_X) + S_\one)\\
		&= \delta_a^{\mathcal{A}} (S_{\mathcal{A}}) \cdot P\left(f^{-1}(S_X)\right) = (\delta_a^{\mathcal{A}} \odot (P\circ f^{-1}))(S_{\mathcal{A}} \times S_X + S_\one) \\
		&= (\lambda_X \circ FTf)(a,P)(S)
\end{align*}
and analogously we obtain
\begin{align*}
	&(TFf \circ \lambda_Y) (\checkmark) (S) =\delta_\checkmark^{\mathcal{A}\times Y + \one}\left((Ff)^{-1} (S)\right) \\
	&\quad=\delta_\checkmark^{\mathcal{A}\times Y + \one}\left(S_{\mathcal{A}}  \times f^{-1}(S_X) + S_\one\right) = \delta_\checkmark^{\mathcal{A}\times X + \one} (S)= (\lambda_X \circ FTf)(\checkmark)(S)\,.
\end{align*}
\item For $(a,x) \in \A \times X$ we calculate
\begin{align*}
	\eta_{FX}(a,x)(S) &= \delta_{(a,x)}^{FX}(S_\mathcal{A} \times S_X + S_\one) = \delta_a^\mathcal{A}(S_\mathcal{A}) \cdot \delta_x^X(S_X)\\
	&= (\delta_a^\mathcal{A} \odot \delta_x^X)(S) =\lambda_X(a,\delta_x^X)(S) = \big(\lambda_X \circ F\eta_X\big)(a,x)(S)
\end{align*}
and also
\begin{align*}	
	\eta_{FX} (\checkmark) = \delta_\checkmark^{FX} = \lambda_X(\checkmark) = \lambda_X\left(F\eta_X(\checkmark)\right) = \big(\lambda_X\circ F\eta_X\big)(\checkmark)\,.
\end{align*}
\item For $(a,P) \in FT^2X$ we calculate 
\begin{align*}
	\left(\lambda_X \circ F\mu_X\right)(a,P)(S) &= \left(\lambda_X\left(a, \mu_X(P)\right)\right)(S) = \left(\delta_a^\mathcal{A}\odot \mu_X(P)\right)(S) \\
	&=\delta_a^\mathcal{A}(S_{\mathcal{A}}) \cdot \mu_X(P)(S_X) = \delta_a^\mathcal{A}(S_{\mathcal{A}}) \cdot \Int{p_{S_X}}[P]
\end{align*}
and 
\begin{align*}
	&\left(\mu_{FX} \circ T\lambda_X\circ  \lambda_{TX}\right)\!(a,P)(S) = \mu_{FX}\left(\left(\delta_a^\mathcal{A} \odot P\right) \circ \lambda^{-1}_X\right)(S) \\
	&\quad = \Int[TFX]{p_S}[{\left(\left(\delta_a^\mathcal{A}\odot P\right) \circ \lambda^{-1}_X\right)}] = \Int[\lambda^{-1}_X(TFX)]{p_S\circ \lambda_X}[\big(\delta_a^\mathcal{A} \odot P\big)] \\
	&\quad = \Int[\set{a} \times TX]{p_S\circ \lambda_X}[\big(\delta_a^\mathcal{A} \odot P\big)] = \Int[P' \in TX]{\big(\delta_a^\mathcal{A} \otimes P')(S)}[P(P')]\\
	&\quad = \Int[P' \in TX]{\delta_a^\mathcal{A}(S_{\mathcal{A}})\cdot P'(S_X)}[P(P')] = \delta_a^\mathcal{A}(S_{\mathcal{A}})\cdot \Int{p_{S_X}}[P]\,.
\end{align*}
Analogously we obtain
\begin{align*}
	\left(\lambda_X \circ F\mu_X \right)(\checkmark) = \lambda_X(\checkmark) = \delta_{\checkmark}^{\mathcal{A}\times X+1}
\end{align*}
and
\begin{align*}
	&\left(\mu_{FX} \circ T\lambda_X\circ \lambda_{TX}\right)(\checkmark)(S) = \mu_{FX}\left(\delta_\checkmark^{\mathcal{A} \times TX+\one} \circ \lambda^{-1}_X\right)(S)\\
	&\quad=\Int[TFX]{p_S}[\left(\delta_\checkmark^{\mathcal{A}\times TX+\one}\circ \lambda^{-1}_X\right) ] = \Int[\lambda^{-1}_X(TFX)]{p_S \circ \lambda_X}[\delta_\checkmark^{\mathcal{A}\times TX+\one}]\\
	&\quad = (p_S \circ \lambda_X)(\checkmark)= \delta_\checkmark^{\mathcal{A}\times X+\one}(S)\,. 
\end{align*}
\end{enumerate}
\qed

\noindent With this result at hand we can finally apply Corollary \ref{cor:main_corollary} to the measurable spaces $\emptyset$, $\Astar$, $\Aomega$, $\Ainfty$, each of which is of course equipped with the $\sigma$-algebra generated by the covering semirings $\S_0$, $\Sstar$, $\Somega$, $\Sinfty$ as defined in Proposition~\ref{prop:semirings_of_words}, to obtain the final coalgebra and the induced trace semantics for PTS as presented in the following theorem. 

\begin{thm}[Final Coalgebra and Trace Semantics for PTS]
Let $(T, \eta, \mu)$ be either the sub-probability monad $(\mathbb{S}, \eta, \mu)$ or the probability monad $(\mathbb{P}, \eta, \mu)$ and $F$ be either $\A \times \Id_\Meas$ or $\A \times \Id_\Meas + \one$. A PTS $(\A, X, \alpha)$ is an $\overline{F}$-coalgebra $(X , \alpha)$ in $\Kl(T)$ and vice versa. In the following table we present the (carriers of) final $\overline{F}$-coalgebras $\left(\Omega, \kappa\right)$ in $\Kl(T)$ for all suitable choices of $T$ and $F$ (depending on the type of the PTS).
\begin{align*}\begin{tabular}{c|c|l|c}
	\hline
	Type $\diamond$~ 	& ~Monad $T$ ~ 	& ~Endofunctor $F$ ~ 			& Carrier $\Omega$ \\\hline
	$0$ 				& $\mathbb{S}$ 	& ~$\A \times \Id_\Meas$ 			& $(\emptyset, \set{\emptyset})$   \\
	$*$ 				& $\mathbb{S}$ 	& ~$\A \times \Id_\Meas + \one$	& $\left(\Astar, \sigalg[\Astar]{\Sstar}\right) $ \\
	$\omega$ 		& $\mathbb{P}$ 	& ~$\A \times \Id_\Meas$			& $\left(\Aomega, \sigalg[\Aomega]{\Somega}\right) $\\
	$\infty$ 			& $\mathbb{P}$ 	& ~$\A \times \Id_\Meas + \one$	& ~$\left(\Ainfty, \sigalg[\Ainfty]{\Sinfty}\right)$\\\hline
\end{tabular}\end{align*}
where for $\diamond \in \set{*, \omega, \infty}$ we have $\kappa = \eta_{F\Omega}\circ \phi$ where $\phi$ is the isomorphism as defined in Proposition~\ref{prop:words_iso} and for $\diamond = \emptyset$ we take $\kappa = \eta_{F\emptyset} \circ \phi$ with $\phi$ being the empty function $\phi \colon \emptyset \to \emptyset$. The unique arrow into the final coalgebra is the map $\tr\colon X \to T\Omega$ given by Definition~\ref{def:trace_premeasure}.
\end{thm}
\proof
For the whole proof we always assume that the combinations of the type $\diamond$ of the PTS, the monad $T$, the endofunctor $F$ and the carrier $(\Omega, \Sigma_\Omega)$ are chosen as presented in the table given in the corollary. Thus e.g. $\diamond = *$ automatically yields $T = \mathbb{S}$, $F = \A \times \Id_\Meas + \one$, $\Omega=\Astar$, $\Sigma_\Omega = \sigalg[\Astar]{\Sstar}$ and we automatically work in the Kleisli category $\Kl(\mathbb{S})$ of the sub-probability monad. The first statement of the theorem is obvious by construction of the transition function $\alpha$. For $\diamond \in \set{*, \omega, \infty}$ we remark that the preconditions of Corollary \ref{cor:main_corollary} are fulfilled and aim at applying this corollary, and especially at evaluating equation \eqref{eq:giry_final_coalgebra_semiring} for the covering semirings $\Sstar, \Somega, \Sinfty$. Let us carry out these calculations in various steps to obtain all the equations of Definition \ref{def:trace_premeasure}. For all $(b,x') \in \A \times X$ we calculate
\begin{align*}
	(\lambda_\Omega \circ F(\tr)) (b,x') =
	\begin{cases} 	
		\delta_b^\A \otimes \tr(x'), & \diamond = \omega\\
		\delta_b^\A \odot \tr(x'), & \diamond \in \set{*, \infty}.
	\end{cases}
\end{align*}
Now suppose $S$ is chosen as $S=\set{au}$, $S=~\cone{\omega}{au}$ or $S=~\cone{\infty}{au}$ respectively for an arbitrary $a \in \A$ and an arbitrary $u \in \Astar$. Then $\phi(S) = \set{a} \times S'$ with $S'=\set{u}$, $S'=~\cone{\omega}{u}$ or $S'=~\cone{\infty}{u}$ respectively and hence we obtain
\begin{align*}
	&(p_{\phi(S)} \circ \lambda_\Omega \circ F(\tr)) (b,x') = \delta_b^\A \otimes \tr(x')(\set{a} \times S') \\
		&\quad = \delta_b^\A(\set{a}) \cdot \tr(x')(S') = \chi_{\set{a} \times X}(b,x') \cdot \tr(x')(S')\,.
\end{align*}
Using this, we evaluate equation \eqref{eq:giry_final_coalgebra_semiring} of Corollary \ref{cor:main_corollary} for these sets and get
\begin{align*}
	\tr(x)(S) = \Int[(b,x') \in \set{a} \times X]{\tr(x')(S')}[\alpha(x)] = \Int[x' \in X]{\tr(x')(S')}[\P{a}{x}{x'}]
\end{align*}
which yields equations \eqref{eq:trace_main_equation} and \eqref{eq:trace_main_equation2} of Definition~\ref{def:trace_premeasure}. For $\diamond \in \set{*,\infty}$ we calculate
\begin{align*}
	(\lambda_\Omega \circ F(\tr)) (\checkmark) = \delta_\checkmark^{\A \times \Omega + \one}
\end{align*}
and conclude that for $z \in \A \times X + \one$ we have $(p_{\phi(\set{\epsilon})} \circ \lambda_\Omega \circ F(\tr)) (z) = 1$ if and only if $z = \checkmark$. Hence evaluating equation \eqref{eq:giry_final_coalgebra_semiring} in this case yields
\begin{align*}
	\tr(x)(\set{\epsilon}) = \Int{p_{\phi(\set{\epsilon})} \circ \lambda_\Omega \circ F(\tr)}[\alpha(x)] =  \Int{\chi_\one}[\alpha(x)] = \alpha(x)(\one)
\end{align*}
which is equation \eqref{eq:trace_emptyword}. For $\diamond \in \set{\omega, \infty}$ we have $\tr(x)(\A^\diamond) = 1$ due to the fact that $\tr(x)$ must be a probability measure. This is already equation \eqref{eq:trace_wholespace} because $\A^\diamond=\epsilon \A^\diamond$. Moreover $\phi(\cone{\diamond}{\epsilon}) = \phi(\Omega) = F\Omega$ and since also $\lambda_\Omega \circ F(\tr)$ must be a probability measure evaluating \eqref{eq:giry_final_coalgebra_semiring} yields the same:
\begin{align*}
	\tr(x)(\cone{\diamond}{\epsilon}) &= \Int{p_{\phi(\cone{\diamond}{\epsilon})} \circ \lambda_\Omega \circ F(\tr)}[\alpha(x)] =\Int{1}[\alpha(x)] = \alpha(x)(FX) = 1\,.
\end{align*}
Finally, for $\diamond=0$ we remark, that the $\Kl(\mathbb{S})$-object $(\emptyset, \set{\emptyset})$ is the unique final object of $\Kl(\mathbb{S})$: Given any $\Kl(\mathbb{S})$-object $(X, \Sigma_X)$, the unique map into the final object is given as $f \colon X \to \mathbb{S}(\emptyset) = \set{(p \colon  \set{\emptyset} \to [0,1], p(\emptyset) = 0)}$ mapping any $x \in X$ to the unique element of that set. Moreover, $(\emptyset, \set{\emptyset})$ together with $\kappa = \eta_{F\emptyset} \circ \phi$, where the map $\phi \colon \emptyset \to \A \times \emptyset$ is the obvious and unique isomorphism $(\emptyset, \powerset{\emptyset}) \cong (\A \times \emptyset, \powerset{\A} \otimes \powerset{\emptyset})$, is a $\overline{F}$-coalgebra and thus final. 

In all cases we have obtained exactly the equations from Definition~\ref{def:trace_premeasure} which by Proposition \ref{prop:trace_premeasure} yield a unique function $\tr\colon X \to T\A^\diamond$. From Proposition \ref{prop:TraceIsKleisli} we know that this function is indeed a Kleisli arrow.\qed

\section{Examples}
\label{sec:advexamples}
In this section we will define and examine two truly continuous probabilistic systems and calculate their trace measures or parts thereof. However, in order to deal with these systems, we first need to provide some additional measure theoretic results and tools. At first, we will explain the \emph{counting measure} on countable sets and also the \emph{Lebesgue measure} as this is ``the'' standard measure on the reals. Afterwards we will take a quick look into the theory of measures with \emph{densities}. With these tools at hand we can finally present the examples. All of the presented results should be contained in any standard textbook on measure and integration theory. We use \cite{Els07} as our primary source for this short summary.

\begin{defi}[Counting Measure]
Let $X$ be a countable set. The \emph{counting measure} on $(X, \powerset{X})$ is the cardinality map 
\begin{align}
	\#\colon \powerset{X} \to \ExtR_+, \quad A \mapsto |A|
\end{align}
assigning to each finite subset of $X$ its number of elements and $\infty$ to each infinite subset of $X$. It is uniquely defined as the extension of the $\sigma$-finite pre-measure on the set of all singletons (and $\emptyset$) which is $1$ on every singleton and $0$ on $\emptyset$.
\end{defi}

\subsection{Completion and the Lebesgue Measure}
The (one-dimensional) \emph{Lebesgue-Borel measure} is the unique measure $\lambda'$ on the reals equipped with the Borel $\sigma$-algebra $\Borel(\R)$ satisfying $\lambda'\left((a,b]\right) = b-a$ for every $a, b \in \R$, $a \leq b$. In order to obtain the \emph{Lebesgue measure}, we will refine both the measure and the set of measurable sets by \emph{completion}. We call a measure space $(X, \Sigma, \mu)$ \emph{complete} if every subset of a $\mu$-null-set (i.e. a measurable set $S \in \Sigma$ such that $\mu(S) = 0$) is measurable (and necessarily also a $\mu$-null-set). For any measure space $(X, \Sigma, \mu)$ there is always a smallest complete measure space $(X, \tilde{\Sigma}, \tilde{\mu})$ such that $\Sigma \subseteq \tilde{\Sigma}$ and $\tilde{\mu}|_\Sigma = \mu$ called the \emph{completion} (\cite[II.~§6]{Els07}). The completion of the Lebesgue-Borel measure yields the \emph{Lebesgue $\sigma$-algebra} $\Lebesgue$ and the \emph{Lebesgue measure}\footnote{This is the second meaning of the symbol $\lambda$. Until here, $\lambda$ was used as symbol for a distributive law.} $\lambda \colon \Lebesgue \to \ExtR$. For the Lebesgue measure we will use the following notation for integrals:
\begin{align*}
	\Int[a][b]{f} := \Int[{[a,b]}]{f}[\lambda]\,.
\end{align*}

\subsection{Densities}
When dealing with measures on arbitrary measurable spaces -- especially in the context of probability measures -- it is sometimes useful to describe them using so-called \emph{densities}. We will give a short introduction into the theory of densities here which is sufficient for understanding the upcoming examples. Given a measurable space $(X, \Sigma_X)$ and measures $\mu, \nu \colon \Sigma_X \to \ExtR_+$ we call a Borel-measurable function $f \colon X \to \ExtR$ satisfying
\begin{align}
	\nu (S) = \Int[S]{f}[\mu]\label{eq:density}
\end{align}
for all measurable sets $S \in \Sigma_X$ a \emph{$\mu$-density of $\nu$}. In that case $\mu(S) = 0$ implies $\nu(S)=0$ for all measurable sets $S \in \Sigma_X$ and we say that $\nu$ is \emph{absolutely continuous} with respect to $\mu$ and write $\nu \ll \mu$. Densities are neither unique nor do they always exist. However, if $\nu$ has two $\mu$-densities $f,g$ then $f = g$ holds $\mu$-almost everywhere, i.e. there is a $\mu$ null set $N \in \Sigma_X$ such that for all $x \in X\setminus N$ we have $f(x) = g(x)$.  Moreover, any such $\mu$-density uniquely defines the measure $\nu$. If $\mu = \lambda$, i.e. $\mu$ is the Lebesgue-measure, and \eqref{eq:density} holds for a measure $\nu$ and a function $f$, then $f$ is called \emph{Lebesgue density} of $\nu$. For our examples we will make use of the following Proposition which can be found e.g. in \cite[IV.2.12 Satz]{Els07}.

\begin{prop}[Integration and Measures with Densities]
Let $(X, \Sigma_X)$ be a measurable space and let $\mu, \nu \colon \Sigma_X \to \R_+$ be measures such that $\nu$ has a $\mu$-density $f$. If $g \colon X \to \R_+$ is $\nu$-integrable, then $\int\!g\,\mathrm{d}\nu = \int\!gf\,\mathrm{d}\mu.$\qed
\end{prop}

\subsection{Examples}
With all the previous results at hand, we can now present our two continuous examples using densities to describe the transition functions.

\begin{exa}
We will first give an informal description of this example as a kind of one-player-game which is played in the closed real interval $[0,1]$. The player, who is in any point $z \in [0,1]$, can jump up and will afterwards touch down on a new position $x \in [0,1]$ which is determined probabilistically. After a jump, the player announces, whether he is left ``$L$'' or right ``$R$'' of his previous position. The total probability of jumping from $z$ to the left is $z$ and the probability of jumping to the right is $1-z$. In both cases, we have a continuous uniform probability distribution. As we are within the set of reals, the probability of hitting a specific point $x_0 \in [0,1]$ is always zero. Let us now continue with the precise definition of our example.
Let $\mathcal{A} := \set{L,R}$. We consider the PTS $(\set{L,R}, [0,1], \alpha)$ where $[0,1]$ is equipped with the Lebesgue $\sigma$-algebra of the reals, restricted to that interval denoted $\Lebesgue([0,1])$. The transition probability function $\alpha \colon [0,1] \to \mathbb{P}([0,1])$ is given as
\begin{align*}
	\alpha(z)(S) = \Int[S]{f_z}[(\# \otimes \lambda)]
\end{align*}
for every $z \in [0,1]$ and all sets $S \in \powerset{\set{L,R}} \otimes \Lebesgue([0,1])$ with the $(\# \otimes \lambda)$-densities
\begin{align*}
	f_z \colon \set{L,R} \times [0,1] \to \R^+, \quad (a,x) \mapsto \chi_{\set{L} \times [0,z]}(a,x) + \chi_{\set{R} \times [z,1]}(a,x)\,.
\end{align*}
We observe that $S \mapsto \P{L}{z}{S}, S \mapsto \P{R}{z}{S} \colon \Lebesgue([0,1]) \to \R^+$ thus have Lebesgue-densities
\begin{align*}
	\P{L}{z}{S} = \Int[S]{\chi_{[0,z]}}[\lambda] = \Int[S]{\chi_{[0,z]}(x)}, \quad \P{R}{z}{S} = \Int[S]{\chi_{[z,1]}}[\lambda] = \Int[S]{\chi_{[z,1]}(x)}\,.
\end{align*}
with the following graphs (in the real plane)
\begin{center}
\begingroup
  \makeatletter
  \providecommand\color[2][]{%
    \GenericError{(gnuplot) \space\space\space\@spaces}{%
      Package color not loaded in conjunction with
      terminal option `colourtext'%
    }{See the gnuplot documentation for explanation.%
    }{Either use 'blacktext' in gnuplot or load the package
      color.sty in LaTeX.}%
    \renewcommand\color[2][]{}%
  }%
  \providecommand\includegraphics[2][]{%
    \GenericError{(gnuplot) \space\space\space\@spaces}{%
      Package graphicx or graphics not loaded%
    }{See the gnuplot documentation for explanation.%
    }{The gnuplot epslatex terminal needs graphicx.sty or graphics.sty.}%
    \renewcommand\includegraphics[2][]{}%
  }%
  \providecommand\rotatebox[2]{#2}%
  \@ifundefined{ifGPcolor}{%
    \newif\ifGPcolor
    \GPcolortrue
  }{}%
  \@ifundefined{ifGPblacktext}{%
    \newif\ifGPblacktext
    \GPblacktexttrue
  }{}%
  \let\gplgaddtomacro\g@addto@macro
  \gdef\gplbacktext{}%
  \gdef\gplfronttext{}%
  \makeatother
  \ifGPblacktext
    \def\colorrgb#1{}%
    \def\colorgray#1{}%
  \else
    \ifGPcolor
      \def\colorrgb#1{\color[rgb]{#1}}%
      \def\colorgray#1{\color[gray]{#1}}%
      \expandafter\def\csname LTw\endcsname{\color{white}}%
      \expandafter\def\csname LTb\endcsname{\color{black}}%
      \expandafter\def\csname LTa\endcsname{\color{black}}%
      \expandafter\def\csname LT0\endcsname{\color[rgb]{1,0,0}}%
      \expandafter\def\csname LT1\endcsname{\color[rgb]{0,1,0}}%
      \expandafter\def\csname LT2\endcsname{\color[rgb]{0,0,1}}%
      \expandafter\def\csname LT3\endcsname{\color[rgb]{1,0,1}}%
      \expandafter\def\csname LT4\endcsname{\color[rgb]{0,1,1}}%
      \expandafter\def\csname LT5\endcsname{\color[rgb]{1,1,0}}%
      \expandafter\def\csname LT6\endcsname{\color[rgb]{0,0,0}}%
      \expandafter\def\csname LT7\endcsname{\color[rgb]{1,0.3,0}}%
      \expandafter\def\csname LT8\endcsname{\color[rgb]{0.5,0.5,0.5}}%
    \else
      \def\colorrgb#1{\color{black}}%
      \def\colorgray#1{\color[gray]{#1}}%
      \expandafter\def\csname LTw\endcsname{\color{white}}%
      \expandafter\def\csname LTb\endcsname{\color{black}}%
      \expandafter\def\csname LTa\endcsname{\color{black}}%
      \expandafter\def\csname LT0\endcsname{\color{black}}%
      \expandafter\def\csname LT1\endcsname{\color{black}}%
      \expandafter\def\csname LT2\endcsname{\color{black}}%
      \expandafter\def\csname LT3\endcsname{\color{black}}%
      \expandafter\def\csname LT4\endcsname{\color{black}}%
      \expandafter\def\csname LT5\endcsname{\color{black}}%
      \expandafter\def\csname LT6\endcsname{\color{black}}%
      \expandafter\def\csname LT7\endcsname{\color{black}}%
      \expandafter\def\csname LT8\endcsname{\color{black}}%
    \fi
  \fi
  \setlength{\unitlength}{0.0500bp}%
  \begin{picture}(5668.00,1700.00)%
    \gplgaddtomacro\gplbacktext{%
      \csname LTb\endcsname%
      \put(330,457){\makebox(0,0)[r]{\strut{}}}%
      \put(330,1098){\makebox(0,0)[r]{\strut{}$1$}}%
      \put(462,174){\makebox(0,0){\strut{}$0$}}%
      \put(1803,174){\makebox(0,0){\strut{}$z$}}%
      \put(4644,174){\makebox(0,0){\strut{}$1$}}%
      \csname LTb\endcsname%
      \put(880,1354){\makebox(0,0)[l]{\strut{}$\chi_{[0,z]}$}}%
      \put(880,713){\makebox(0,0)[l]{\strut{}$\chi_{[z,1]}$}}%
      \put(2887,1354){\makebox(0,0)[l]{\strut{}$\chi_{[z,1]}$}}%
      \put(2887,713){\makebox(0,0)[l]{\strut{}$\chi_{[0,z]}$}}%
    }%
    \gplgaddtomacro\gplfronttext{%
    }%
    \gplbacktext
    \put(0,0){\includegraphics{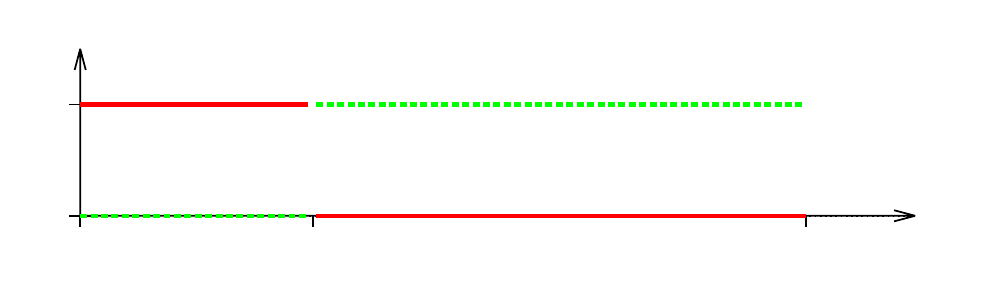}}%
    \gplfronttext
  \end{picture}%
\endgroup

 \end{center}
Evaluating these measures on $[0,1]$ yields
\begin{align*}
	\P{L}{z}{[0,1]} = \int_0^z\!1\,\mathrm{d}x = z, \quad \P{R}{z}{[0,1]} = \int_z^1\! 1\,\mathrm{d}x = 1-z\,.
\end{align*}
With these preparations at hand we calculate the trace measure on some cones.
\begin{align*}
	\tr(z)(\cone{\omega}{\epsilon}) &= 1\\
	\tr(z)(\cone{\omega}{L}) &= \Int[{[0,1]}]{1}[\P{L}{z}{z'}] = \P{L}{z}{[0,1]} = z\\
	\tr(z)(\cone{\omega}{R}) &= \Int[{[0,1]}]{1}[\P{R}{z}{z'}] = \P{R}{z}{[0,1]} = 1-z\\
	\tr(z)(\cone{\omega}{LL}) &= \Int[{[0,1]}]{x}[\P{L}{z}{x}] = \Int[0][1]{x \cdot \chi_{[0,z]}(x)} = \Int[0][z]{x} = \left[\frac{1}{2} x^2\right]_0^z = \frac{1}{2}z^2\\
	\tr(z)(\cone{\omega}{LR}) &= \Int[{[0,1]}]{1-x}[\P{L}{z}{x}] = \Int[0][z]{(1-x)} = \left[x-\frac{1}{2} x^2\right]_0^z = z- \frac{1}{2}z^2\\
	\tr(z)(\cone{\omega}{RL}) &= \Int[{[0,1]}]{x}[\P{R}{z}{x}] = \Int[0][1]{x \cdot \chi_{[z,1]}(x)} = \Int[z][1]{x} = \left[\frac{1}{2} x^2\right]_z^1 = \frac{1}{2} -  \frac{1}{2} z^2\\
	\tr(z)(\cone{\omega}{RR}) &= \Int[{[0,1]}]{1-x}[\P{R}{z}{x}] = \Int[z][1]{(1-x)} = \left[x-\frac{1}{2} x^2\right]_z^1 = \frac{1}{2} -  z + \frac{1}{2} z^2
\end{align*}
Thus for any word $u \in \Astar$ of length $n$ there is a polynomial $p_u \in \R[Z]$ in one variable $Z$ with degree $\mathop{deg}(p_u) = n$. Evaluating this polynomial for an arbitrary $z \in [0,1]$ yields the value of the trace measure $\tr(z)$ on the cone $\cone{\omega}{u}$ generated by $u$, i.e. $\tr(z)(\cone{\omega}{u}) = p_u(z)$.
\end{exa}

While the previous example provides some understanding on how to describe a continuous PTS and also on how to calculate its trace measure, we are interested in trace equivalence. The second example will thus be a system which is trace equivalent to a finite state system.

\begin{exa}
As before, we will give an informal description as a kind of one-player-game first. There is exactly one player, who starts in any point $z \in \R$, jumps up and touches down somewhere on the real line announcing whether he is left ``$L$'' or right ``$R$'' of his previous position or has landed back on his previous position ``$N$''. The probability of landing is initially given via a normal distribution centered on the original position $z$. Thus, the probability of landing in close proximity of $z$, i.e. in the interval $[z-\epsilon, z + \epsilon]$, is high for sufficiently big $\epsilon \in \R_+\setminus\set{0}$ whereas the probability of landing far away, i.e. outside of that interval, is negligible. The player has a finite amount of energy and each jump drains that energy so that after finitely many jumps he will not be able to jump again resulting in an infinite series of ``$N$'' messages. Before that the energy level determines the likelihood of his jump width, i.e. the standard deviation of the normal distributions. Now let us give a formal description of such a system. Recall that the density function of the normal distribution with expected value\footnote{This is the third meaning of $\mu$. Until here, $\mu$ was used as symbol for a measure and also as a symbol for the multiplication natural transformation of a monad.} $\mu \in \R$ and standard deviation $\sigma \in \R^+ \setminus\set{0}$ is the Gaussian function
\begin{align*}
	\phi_{\mu, \sigma}\colon \R \to \R^+, \quad \phi_{\mu, \sigma}(x) =  \frac{1}{\sigma \sqrt{2\pi}} \cdot \exp\left(-\frac{1}{2}\left(\frac{x-\mu}{\sigma}\right)^2\right)
\end{align*}
with the following graph (in the real plane), often called the ``bell curve".
\begin{center}
\begingroup
  \makeatletter
  \providecommand\color[2][]{%
    \GenericError{(gnuplot) \space\space\space\@spaces}{%
      Package color not loaded in conjunction with
      terminal option `colourtext'%
    }{See the gnuplot documentation for explanation.%
    }{Either use 'blacktext' in gnuplot or load the package
      color.sty in LaTeX.}%
    \renewcommand\color[2][]{}%
  }%
  \providecommand\includegraphics[2][]{%
    \GenericError{(gnuplot) \space\space\space\@spaces}{%
      Package graphicx or graphics not loaded%
    }{See the gnuplot documentation for explanation.%
    }{The gnuplot epslatex terminal needs graphicx.sty or graphics.sty.}%
    \renewcommand\includegraphics[2][]{}%
  }%
  \providecommand\rotatebox[2]{#2}%
  \@ifundefined{ifGPcolor}{%
    \newif\ifGPcolor
    \GPcolortrue
  }{}%
  \@ifundefined{ifGPblacktext}{%
    \newif\ifGPblacktext
    \GPblacktexttrue
  }{}%
  \let\gplgaddtomacro\g@addto@macro
  \gdef\gplbacktext{}%
  \gdef\gplfronttext{}%
  \makeatother
  \ifGPblacktext
    \def\colorrgb#1{}%
    \def\colorgray#1{}%
  \else
    \ifGPcolor
      \def\colorrgb#1{\color[rgb]{#1}}%
      \def\colorgray#1{\color[gray]{#1}}%
      \expandafter\def\csname LTw\endcsname{\color{white}}%
      \expandafter\def\csname LTb\endcsname{\color{black}}%
      \expandafter\def\csname LTa\endcsname{\color{black}}%
      \expandafter\def\csname LT0\endcsname{\color[rgb]{1,0,0}}%
      \expandafter\def\csname LT1\endcsname{\color[rgb]{0,1,0}}%
      \expandafter\def\csname LT2\endcsname{\color[rgb]{0,0,1}}%
      \expandafter\def\csname LT3\endcsname{\color[rgb]{1,0,1}}%
      \expandafter\def\csname LT4\endcsname{\color[rgb]{0,1,1}}%
      \expandafter\def\csname LT5\endcsname{\color[rgb]{1,1,0}}%
      \expandafter\def\csname LT6\endcsname{\color[rgb]{0,0,0}}%
      \expandafter\def\csname LT7\endcsname{\color[rgb]{1,0.3,0}}%
      \expandafter\def\csname LT8\endcsname{\color[rgb]{0.5,0.5,0.5}}%
    \else
      \def\colorrgb#1{\color{black}}%
      \def\colorgray#1{\color[gray]{#1}}%
      \expandafter\def\csname LTw\endcsname{\color{white}}%
      \expandafter\def\csname LTb\endcsname{\color{black}}%
      \expandafter\def\csname LTa\endcsname{\color{black}}%
      \expandafter\def\csname LT0\endcsname{\color{black}}%
      \expandafter\def\csname LT1\endcsname{\color{black}}%
      \expandafter\def\csname LT2\endcsname{\color{black}}%
      \expandafter\def\csname LT3\endcsname{\color{black}}%
      \expandafter\def\csname LT4\endcsname{\color{black}}%
      \expandafter\def\csname LT5\endcsname{\color{black}}%
      \expandafter\def\csname LT6\endcsname{\color{black}}%
      \expandafter\def\csname LT7\endcsname{\color{black}}%
      \expandafter\def\csname LT8\endcsname{\color{black}}%
    \fi
  \fi
  \setlength{\unitlength}{0.0500bp}%
  \begin{picture}(5668.00,1700.00)%
    \gplgaddtomacro\gplbacktext{%
      \csname LTb\endcsname%
      \put(2801,223){\makebox(0,0){\strut{}$z$}}%
      \csname LTb\endcsname%
      \put(3233,1234){\makebox(0,0)[l]{\strut{}$\phi$}}%
    }%
    \gplgaddtomacro\gplfronttext{%
    }%
    \gplbacktext
    \put(0,0){\includegraphics{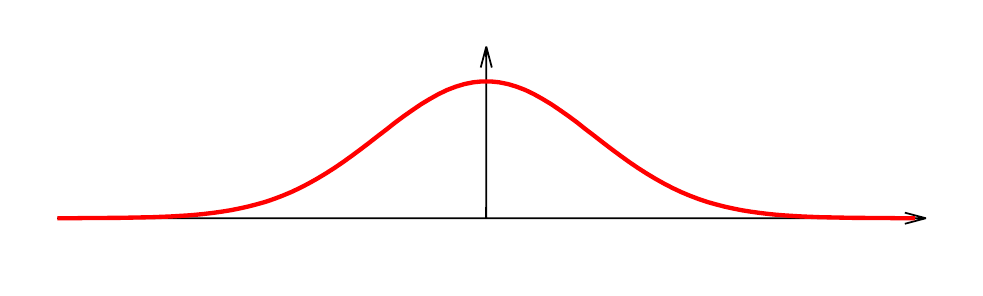}}%
    \gplfronttext
  \end{picture}%
\endgroup

\end{center}
Let now the finite ``energy level'' or ``time horizon'' (which is the maximal number of jumps) $T \in \N$, $T \geq 2$ be given. We consider the PTS with alphabet $\mathcal{A} := \set{L,N,R}$, state space $(\N_0 \times \R, \powerset{\N_0} \otimes \Lebesgue)$ and transition probability function $\alpha\colon \N_0 \times \R \to \Prob{\mathcal{A} \times \N_0 \times \R}$ which we define in two steps. For all $(t,z) \in \N_0 \times \R$ with $t < T$ and all measurable sets $S \in \powerset{\mathcal{A}} \otimes \powerset{\N_0} \otimes \Lebesgue$ we set
\begin{align*}
	\alpha(t,z)(S) := \Int[S]{f_{(t,z)}}[(\# \otimes \# \otimes \lambda)] 
\end{align*}
where the $(\#\otimes\#\otimes\lambda)$-density $f_{(t,z)}$ is
\begin{align*}
	f_{(t,z)}\colon \mathcal{A} \times \N_0 \times \R \to \R^+, (a, t', x) \mapsto
		\begin{cases}
			\chi_{(-\infty,z]}(x) \cdot \phi_{z, 1/(t+1)}(x), & a = L \wedge t' = t+1\\
			\chi_{[z, +\infty)}(x) \cdot \phi_{z, 1/(t+1)}(x), & a = R \wedge t' = t+1\\
			0, & \text{else.}
		\end{cases}
\end{align*}
Thus in the first two cases the density is the left (or right) half of the Gaussian density function with expected value $\mu = z$ and standard deviation $\sigma = 1/(t+1)$ and the constant zero function in all other cases. For the remaining $(t,z) \in \N_0 \times \R$ with $t \geq T$ we define the transition probability function to be 
\begin{align*}
	\alpha(t,z):= \delta_{(N, t+1, z)}^{\mathcal{A} \times \N_0 \times \R}\,.
\end{align*}
We observe that for $(t,z) \in \N_0 \times \R$ with $t < T$ we have $\P{N}{(t,z)}{\N_0 \times \R} = 0$ and 
\begin{align*}
	\P{L}{(t,z)}{\N_0 \times \R} = \Int[-\infty][z]{\phi_{z,1/(t+1)}(x)} = \frac{1}{2} =  \Int[z][\infty]{\phi_{z,1/(t+1)}(x)}  = \P{R}{(t,z)}{\N_0\times \R}.
\end{align*}
For $t \geq T$ we have $\P{N}{(t,z)}{\N_0 \times \R}  = 1$ and $\P{L}{(t,z)}{\N_0 \times \R} = \P{R}{(t,z)}{\N_0 \times \R} = 0$. When we combine these results we obtain the trace measure.
For $t <T$ we get
\begin{align*}
	\tr(t,z) = \sum\limits_{u \in \set{L,R}^{T-t}} \left(\frac{1}{2}\right)^{T-t} \!\cdot \delta_{uN^\omega}^{\mathcal{A}^\omega}
\end{align*}
and for $t \geq T$ the trace measure is $\tr(t,z) = \delta_{N^\omega}^{\mathcal{A}^\omega}$. Obviously the trace measure does not depend on $z$, i.e. $\tr(t,z_1) = \tr(t, z_2)$ for all $t \in \N$ and all $z_1, z_2 \in \R$. Moreover, there is a simple finite state system which is trace equivalent to this system. The finite system has the same alphabet $\A$, its state space is $(\set{0,\hdots,T},\powerset{\set{0, \hdots, T}})$, and the transition function $\alpha\colon \set{0,\hdots,T} \to \Prob{\A\, \times \set{0,\hdots, T}}$ is given as follows
\begin{center}\begin{tikzpicture}[node distance=1.8 and 2.5, on grid, shorten >=1pt, >=stealth', semithick]
	\begin{scope}[state, inner sep=2pt, minimum size=32pt]
		\draw node [draw] (q0) {$0$};
		\draw node [draw, right=of q0] (q1) {$1$}; 
		\draw node [draw, right=of q1] (q2) {$2$};
		\draw node [draw, right=of q2] (q3) {$T-1$};
		\draw node [draw, right=of q3] (q4) {$T$};
	\end{scope}
	\begin{scope}[->]
		\draw (q0) edge[bend left] node[above] {$L, 1/2$} (q1);
		\draw (q0) edge[bend right] node[below] {$R, 1/2$} (q1);
		\draw (q1) edge[bend left] node[above] {$L, 1/2$} (q2);
		\draw (q1) edge[bend right] node[below] {$R, 1/2$} (q2);
		\draw (q3) edge[bend left] node[above] {$L, 1/2$} (q4);
		\draw (q3) edge[bend right] node[below] {$R, 1/2$} (q4);
		\draw (q4) edge[loop right] node[right] {$N, 1$} (q4);
	\end{scope}		
	\draw (q2) edge[dashed] node {} (q3);
\end{tikzpicture}\end{center}
i.e. for $t < T$ we define
\begin{align*}
	\alpha(t) = \frac{1}{2} \cdot  \left(\delta_{(L, t+1)}^{\A \times \set{0,\hdots,T}} + \delta_{(R, t+1)}^{\A \times \set{0,\hdots,T}}\right)
\end{align*}
and for $t = T$ we define $\alpha(t) = \delta_{(N,T)}^{\A \times \set{0,\hdots,T}}$. 
\end{exa}

\section{Conclusion, Related and Future Work}
We have shown how to obtain coalgebraic trace semantics for generative probabilistic transition systems in a general measure-theoretic setting, thereby allowing uncountable state spaces and infinite trace semantics. Especially we have presented final coalgebras for four different types of probabilistic systems.

There is a huge body of work on Markov processes and probabilistic transition systems, but only part of it deals with behavioral equivalences, as in our setting. Even when the focus is on behavioral equivalences, so far usually bisimilarity and related equivalences have been studied (see for instance \cite{larsenskou89}), neglecting the very natural notion of trace equivalence. Furthermore many papers restrict to countable state spaces and discrete probability theory.

Our work is clearly inspired by \cite{hasuo}, which presents the idea to obtain trace equivalence by considering coalgebras in suitable Kleisli categories, generalizing their instantiation of generative probabilistic systems to a general measure-theoretic setting and considering new types of systems. Different from the route we took in this paper, another option might have been to extend the general theorem (Theorem~3.3) of \cite{hasuo}. The theorem gives sufficient conditions under which a final coalgebra in a Kleisli category coincides with an initial algebra in the underlying category $\mathbf{Set}$. This theorem is given for Kleisli categories over $\mathbf{Set}$ and requires that the Kleisli category is $\mathbf{Cppo}$-enriched, i.e., each homset carries a complete partial order with bottom and some additional conditions hold.  This theorem is non-trivial to generalize. First, it would be necessary to extend it to $\mathbf{Meas}$ and second -- and even more importantly -- the requirement of the Kleisli category being $\mathbf{Cppo}$-enriched is quite restrictive. For the case of the sub-probability monad a bottom elements exist (the arrow which maps everything to the constant $0$-measure), but this is not the case for the probability monad, which is the more challenging part, giving rise to infinite words. Hence we would require a different approach, which can also be seen by the fact that in the case of the probability monad the final coalgebra is \emph{not} the initial algebra in $\mathbf{Meas}$.

The study of probabilistic systems using coalgebra is not a new approach. An extensive survey on the coalgebraic treatment of these systems can be found in \cite{Sokolova20115095} including an overview of various different types of transition systems containing probabilistic effects alongside user-input, non-determinism and termination, extensions that we did not consider in this paper (apart from termination).

A thorough consideration of coalgebras and especially theorems guaranteeing the existence of final coalgebras for certain functors on $\Meas$ is given in \cite{viglizzofinal} but since all these are coalgebras in $\Meas$ and not in the Kleisli category over a suitable monad, the obtained behavioral equivalence is probabilistic Larsen-Skou \cite{larsenskou89} bisimilarity instead of trace equivalence and the results do not directly apply to our setting.

Also, in \cite{doberkat2007stochastic} and \cite{Pan09} a very thorough and general overview of properties of labelled Markov processes including the treatment of and the evaluation of temporal logics on probabilistic systems is given. However,  the authors do not explicitly cover a coalgebraic notion of trace semantics.

Infinite traces in a general coalgebraic setting have already been studied in \cite{Cirstea:2010:GIT:1841982.1842047}. However, this generic theory, once applied to probabilistic systems, is restricted to coalgebras with countable carrier while our setting, which is undoubtedly specific and covers only certain functors and branching types, allows arbitrary carriers for coalgebras of probabilistic systems.

As future work we plan to apply the minimization algorithm introduced in \cite{abhkms:coalgebra-min-det} and adapt it to this general setting, by working out the notion of canonical representatives for probabilistic transition system. We are especially interested in comparing this to the canonical representatives for weak and strong bisimilarity presented recently in \cite{eisentrautetal2013}.

Furthermore we plan to define and study a notion of probabilistic trace distance, similar to the distance measure (for bisimilarity) considered in \cite{bw:behavioural-distances,bw:behavioural-pseudometric}. We are also interested in algorithms for calculating this distance, perhaps similar to what has been proposed in \cite{chen} for probabilistic bisimilarity or the more recent on-the-fly algorithm presented in
\cite{baccietal2013}.

\section*{Acknowledgement}
We would like to thank Paolo Baldan, Filippo Bonchi, Mathias Hülsbusch, Sebastian Küpper and Alexandra Silva for discussing this topic with us and giving us some valuable hints. Moreover, we are grateful for the detailed feedback from our reviewers of both, the conference version, \cite{KK12a}, of this paper and also of the version at hand.

\bibliographystyle{alpha}
\bibliography{pts}

\newcommand{\etalchar}[1]{$^{#1}$}
\begin{thebibliography}{vGSST95}

\bibitem[ABH{\etalchar{+}}12]{abhkms:coalgebra-min-det}
Ji\v{r}\'{\i} Ad\'{a}mek, Filippo Bonchi, Mathias H\"{u}lsbusch, Barbara
  K\"{o}nig, Stefan Milius, and Alexandra Silva.
\newblock A coalgebraic perspective on minimization and determinization.
\newblock In Lars Birkedal, editor, {\em Foundations of Software Science and
  Computational Structures}, volume 7213 of {\em Lecture Notes in Computer
  Science}, pages 58--73. Springer, 2012.

\bibitem[Ash72]{ash}
Robert~B. Ash.
\newblock {\em Real Analysis and Probability}.
\newblock Probability and Mathematical Statistics -- A Series of Monographs and
  Textbooks. Academic Press, 111 Fifth Avenue, New York, New York, 1972.

\bibitem[BBLM13]{baccietal2013}
Giorgio Bacci, Giovanni Bacci, Kim~G. Larsen, and Radu Mardare.
\newblock On-the-fly exact computation of bisimilarity distances.
\newblock In Nir Piterman and Scott~A. Smolka, editors, {\em Tools and
  Algorithms for the Construction and Analysis of Systems}, volume 7795 of {\em
  Lecture Notes in Computer Science}, pages 1--15. Springer Berlin Heidelberg,
  2013.

\bibitem[BK08]{baierkatoen2008}
Christel Baier and Joost-Pieter Katoen.
\newblock {\em Principles of Model Checking}.
\newblock The MIT Press, 2008.

\bibitem[C{\^\i}r10]{Cirstea:2010:GIT:1841982.1842047}
Corina C{\^\i}rstea.
\newblock Generic infinite traces and path-based coalgebraic temporal logics.
\newblock {\em Electronic Notes in Theoretical Computer Science},
  264(2):83--103, August 2010.

\bibitem[CvBW12]{chen}
Di~Chen, Franck van Breugel, and James Worrell.
\newblock On the complexity of computing probabilistic bisimilarity.
\newblock In Lars Birkedal, editor, {\em Foundations of Software Science and
  Computational Structures}, volume 7213 of {\em Lecture Notes in Computer
  Science}, pages 437--451. Springer, 2012.

\bibitem[Dob07a]{doberkat2007stochastic}
E.E. Doberkat.
\newblock {\em Stochastic relations: foundations for Markov transition
  systems}.
\newblock Chapman \& Hall/CRC studies in informatics series. Chapman \&
  Hall/CRC, 2007.

\bibitem[Dob07b]{Dob07b}
Ernst-Erich Doberkat.
\newblock Kleisli morphisms and randomized congruences for the giry monad.
\newblock {\em Journal of Pure and Applied Algebra}, 211(3):638 -- 664, 2007.

\bibitem[Dud89]{Dud89}
Richard~M. Dudley.
\newblock {\em Real Analysis and Probability}.
\newblock Wadsworth \& Brooks/Cole Publishing Company, Pacific Groove,
  California 93950, 1989.

\bibitem[EHS{\etalchar{+}}13]{eisentrautetal2013}
Christian Eisentraut, Holger Hermanns, Johann Schuster, Andrea Turrini, and
  Lijun Zhang.
\newblock The quest for minimal quotients for probabilistic automata.
\newblock In Nir Piterman and Scott~A. Smolka, editors, {\em Tools and
  Algorithms for the Construction and Analysis of Systems}, volume 7795 of {\em
  Lecture Notes in Computer Science}, pages 16--31. Springer Berlin Heidelberg,
  2013.

\bibitem[Els07]{Els07}
J{\"u}rgen Elstrodt.
\newblock {\em Ma{\ss}- und Integrationstheorie}.
\newblock Springer-Lehrbuch. Springer Berlin Heidelberg, 5 edition, 2007.

\bibitem[Gir82]{Gir82}
Mich{\`e}le Giry.
\newblock A categorical approach to probability theory.
\newblock In B.~Banaschewski, editor, {\em Categorical Aspects of Topology and
  Analysis}, volume 915 of {\em Lecture Notes in Mathematics}, pages 68--85.
  Springer Berlin Heidelberg, 1982.

\bibitem[HJS06]{Hasuo06generictrace}
Ichiro Hasuo, Bart Jacobs, and Ana Sokolova.
\newblock Generic trace theory.
\newblock In {\em International Workshop on Coalgebraic Methods in Computer
  Science}, volume 164 of {\em Electronic Notes in Theoretical Computer
  Science}, pages 47--65. Elsevier, 2006.

\bibitem[HJS07]{hasuo}
Ichiro Hasuo, Bart Jacobs, and Ana Sokolova.
\newblock Generic trace semantics via coinduction.
\newblock {\em Logical Methods in Computer Science}, 3 (4:11):1--36, November
  2007.

\bibitem[JR97]{jacobs}
Bart Jacobs and Jan Rutten.
\newblock A tutorial on (co)algebras and (co)induction.
\newblock {\em Bulletin of the European Association for Theoretical Computer
  Science}, 62:222--259, 1997.

\bibitem[Ker11]{kerstan}
Henning Kerstan.
\newblock Trace semantics for probabilistic transition systems - a coalgebraic
  approach.
\newblock Diploma thesis, Universit{\"a}t Duisburg-Essen, September 2011.

\bibitem[KK12a]{KK12a}
Henning Kerstan and Barbara K{\"o}nig.
\newblock Coalgebraic trace semantics for probabilistic transition systems
  based on measure theory.
\newblock In Maciej Koutny and Irek Ulidowski, editors, {\em CONCUR 2012 --
  Concurrency Theory}, volume 7454 of {\em Lecture Notes in Computer Science},
  pages 410--424. Springer Berlin Heidelberg, 2012.

\bibitem[KK12b]{KK12TR}
Henning Kerstan and Barbara K\"{o}nig.
\newblock Coalgebraic trace semantics for probabilistic transition systems
  based on measure theory.
\newblock Technical Report 2012-02, Abteilung f{\"ur} Informatik und Angewandte
  Kognitionswissenschaft, Universit{\"a}t Duisburg-Essen, 2012.

\bibitem[K{\"o}n36]{koenigd}
D\'{e}nes K{\"o}nig.
\newblock {\em Theorie der endlichen und unendlichen {G}raphen}.
\newblock Chelsea Publishing Company New York, N.Y., 1936.

\bibitem[LS89]{larsenskou89}
Kim~G. Larsen and Arne Skou.
\newblock Bisimulation through probabilistic testing.
\newblock In {\em in ``Conference Record of the 16th ACM Symposium on
  Principles of Programming Languages (POPL}, pages 344--352, 1989.

\bibitem[Mul94]{mulry-lifting}
Philip~S. Mulry.
\newblock Lifting theorems for {K}leisli categories.
\newblock In Stephen Brookes, Michael Main, Austin Melton, Michael Mislove, and
  David Schmidt, editors, {\em Mathematical Foundations of Programming
  Semantics}, volume 802 of {\em Lecture Notes in Computer Science}, pages
  304--319. Springer Berlin Heidelberg, 1994.

\bibitem[Pan09]{Pan09}
Prakash Panangaden.
\newblock {\em Labelled Markov Processes}.
\newblock Imperial College Press, 2009.

\bibitem[Rut00]{r:universal-coalgebra}
J.J.M.M. Rutten.
\newblock Universal coalgebra: a theory of systems.
\newblock {\em Theoretical Computer Science}, 249:3--80, 2000.

\bibitem[Sok05]{s:coalg-ps-phd}
Ana Sokolova.
\newblock {\em Coalgebraic Analysis of Probabilistic Systems}.
\newblock PhD thesis, Technische Universiteit Eindhoven, 2005.

\bibitem[Sok11]{Sokolova20115095}
Ana Sokolova.
\newblock Probabilistic systems coalgebraically: A survey.
\newblock {\em Theoretical Computer Science}, 412(38):5095--5110, 2011.
\newblock CMCS Tenth Anniversary Meeting.

\bibitem[vBW05a]{bw:behavioural-distances}
Franck van Breugel and James Worrell.
\newblock Approximating and computing behavioural distances in probabilistic
  transition systems.
\newblock {\em Theoretical Computer Science}, 360:373--385, 2005.

\bibitem[vBW05b]{bw:behavioural-pseudometric}
Franck van Breugel and James Worrell.
\newblock A behavioural pseudometric for probabilistic transition systems.
\newblock {\em Theoretical Computer Science}, 331:115--142, 2005.

\bibitem[vGSST95]{glabbeek}
Rob van Glabbeek, Scott~A. Smolka, Bernhardt Steffen, and Chris M.~N. Tofts.
\newblock Reactive, generative and stratified models of probabilistic
  processes.
\newblock {\em Information and Computation}, 121:59--80, 1995.

\bibitem[Vig05]{viglizzofinal}
Ignacio Viglizzo.
\newblock Final sequences and final coalgebras for measurable spaces.
\newblock In Jos{\'e} Fiadeiro, Neil Harman, Markus Roggenbach, and Jan Rutten,
  editors, {\em Algebra and Coalgebra in Computer Science}, volume 3629 of {\em
  Lecture Notes in Computer Science}, pages 395--407. Springer, 2005.

\end{thebibliography}

\end{document}